\documentclass[final,5p,times,twocolumn,authoryear]{elsarticle}
\usepackage{graphicx}
\usepackage{subcaption}
\usepackage{booktabs}
\usepackage{multirow}
\usepackage{tabularx} 
\usepackage{stfloats}

\usepackage{amssymb}
\usepackage{amsmath}
\usepackage{hyperref}
\usepackage{lipsum}

\usepackage{listings}
\usepackage{xcolor}

\usepackage{lineno}

\journal{Astronomy $\&$ Computing}

\begin{document}
\begin{frontmatter}


\title{The Ratan Active Region Patches (RARPs) Database: A New Database of Solar Active Region Radio Signatures from the RATAN-600 Telescope}

\author[first]{Maxim Korelov}
\ead{mskorelov@gmail.com}

\author[first]{Irina Knyazeva}
\ead{iknyazeva@gmail.com}

\author[first,third]{Evgenii Kurochkin}
\ead{79046155404@yandex.ru}

\author[first]{Nikolay Makarenko}

\author[second]{Denis Derkach}

\affiliation[first]{organization={Central (Pulkovo) Astronomical Observatory at RAS },
            addressline={Pulkovskoye Shosse 65/1},
            postcode={196140},
            city={Saint-Petersburg},
            country={Russia}}
\affiliation[second]{organization={HSE University},
                    addressline={Myasnitskaya Street 20},
                    postcode={101000},
                    city={Moscow},
                    country={Russia}}
\affiliation[third]{organization={Saint-Petersburg Branch of Special Astrophysical Observatory at RAS},
                    addressline={Pulkovskoye Shosse 65/1, AY},
                    postcode={196140},
                    city={Saint-Petersburg},
                    country={Russia}}

\begin{abstract}
Solar flares and coronal mass ejections, originating from solar active regions (ARs), are the primary drivers of space weather and can disrupt technological systems. Forecasting efforts heavily rely on photospheric magnetic field data from the Space-weather HMI Active Region Patch (SHARPs) data products. However, the crucial energy release occurs higher in the solar corona. Radio observations from instruments like the RATAN-600 telescope directly probe this region, but their scientific use has been hindered by a lack of standardized and accessible data products.
To address this gap, we have developed the Ratan Active Region Patches (RARPs) database, a new public resource of multi-frequency radio spectra for solar ARs. Generated using RATANSunPy software, RARPs provides the first standardized radio counterpart to magnetic field archives. The database contains over 160,000 calibrated AR observations from 2009 to 2025, each including 3-18 GHz spectra and rich metadata.
We demonstrate the scientific utility of this database by using machine learning to forecast solar flares. The radio spectra are first compressed into low-dimensional embedded features using an autoencoder, which are then used as predictors in baseline logistic regression classifiers. We compare the predictive power of these embedded RARPs features with that of the 18 SHARPs magnetic field parameters provided in the SHARPs data product headers. Our results show that while SHARPs data provides superior flare discrimination, the radio signatures in RARPs possess clear predictive potential and, for M-class and above flares, yield lower Brier Scores and positive Brier Skill Scores relative to SHARPs, indicating more accurate probabilistic forecasts for these events. This establishes radio data as a valuable and complementary information source.
The RARPs database significantly lowers the barrier for researchers to incorporate radio diagnostics into their work. By making these coronal signatures readily available, our work enables new multi-wavelength investigations into the physics of ARs and strengthens the foundation for developing more accurate space weather forecasting models.

\end{abstract}


\end{frontmatter}


\setlength{\parskip}{0.25em}

\section{Introduction}
\label{introduction}

The evolution of solar ARs, from emergence to decay, is a fundamental problem in solar physics. 
ARs are shaped by magnetohydrodynamic (MHD) processes that govern flux emergence, magnetic reconnection, and plasma instabilities. 
They provide natural laboratories for testing theories of magnetic field generation and energy release \citep{hewett2008multiscale}. 
Observations of ARs across multiple wavelengths are used to test and refine models of the solar dynamo and magnetic cycle  \citep{van2015evolution}. While significant progress has been made in the AR description, many fundamental questions about their formation and stability remain. Answering these questions requires not only improvements in numerical simulations but also access to large, comprehensive observational datasets to verify theoretical models \citep{toriumi2019flare}.

Beyond their theoretical importance, ARs are the primary source of solar flares and coronal mass ejections (CMEs), the main drivers of space weather. These events release vast amounts of magnetic energy, accelerate particles into interplanetary space,  with consequences for satellites, aviation, communications, and ground-based infrastructure \citep{gallagher2002active, bloomfield2012toward}. Improving forecasts of solar eruptions is therefore both a scientific challenge and a practical necessity. Present forecasting approaches rely heavily on photospheric magnetic field measurements. Databases such as the Space-weather HMI Active Region Patches (SHARPs) and the SOHO/MDI Active Region Patches (SMARPs), derived from SOHO/MDI and SDO/HMI, provide standardized, analysis-ready magnetogram cutouts that have transformed AR research and enabled machine-learning applications in forecasting \citep{bobra2021smarps, sinha2022comparative, knyazeva2019deep}. However, the main energy release occurs higher in the atmosphere, in the chromosphere and corona where conditions are more dynamic, but these measurements are rarely exploited in operational flare-prediction pipelines, and the reliance on spaceborne instruments alone offers limited operational redundancy in the face of instrument degradations, spacecraft anomalies, or telemetry interruptions.

Observations in the radio domain offer a more direct diagnostic of higher atmospheric layers, where the main energy release of flares occurs. Radio emissions are sensitive to pre-eruptive changes in coronal magnetic fields and plasma conditions \citep{peterova2021increased}, and forecasting methods based on long-term spectropolarimetric observations from the RATAN-600 radio telescope have shown significant promise \citep{bogod2018method}. These approaches, often leveraging criteria based on the ratio of radio fluxes at different microwave frequencies \citep{tanaka1975microwave}, can detect specific signatures in polarized radio emission that are indicative of flare release, making radio data a powerful and complementary tool to traditional photospheric measurements. The RATAN-600 radio telescope has played an important role in this field for more than four decades: its high sensitivity, wide spectral coverage, and daily solar patrols have produced a unique archive of radio spectra from the mid-1970s to the present, and as a ground-based facility it provides operational redundancy to spaceborne measurements in space-weather monitoring chains. Long-term RATAN-600 observations have revealed numerous correlations between radio flux evolution, polarization properties, and subsequent eruptive activity, underscoring the potential of radio diagnostics for forecasting, but exploiting this archive for large-sample statistical studies and operational applications requires data products that are easy to access, interpret, and combine with other solar observables.

\textit{Motivation and Goal}

Most existing AR databases and forecasting pipelines emphasize photospheric magnetic fields, even though pre-eruptive processes and flare energy release occur in the chromosphere and corona. This photosphere-centric focus limits the ability to test whether explicitly coronal diagnostics, such as radio spectra, can improve models of AR evolution and flare occurrence. A central objective of this work is therefore to bring coronal-sensitive radio measurements onto the same footing as established photospheric predictors and to quantify their contribution to flare forecasting.

Long-term solar radio observations from RATAN-600 are under-used by the broader community. The data are not organized by active region, and lack standardized, analysis-ready products with uniform calibration and metadata. This makes it difficult to perform reproducible, multi-wavelength studies, to cross-match radio observations with magnetic field archives, and to benchmark radio-based predictors against SHARPs and related products.

The goal of this study is to make coronal-sensitive radio diagnostics readily accessible for AR evolution studies and space-weather applications. To this end, we introduce the Ratan Active Region Patches (RARPs) database: standardized, AR-centered, multi-frequency radio spectra cutouts generated with the \texttt{RATANSunPy} software framework \citep{knyazeva2025ratansunpy}, consistently calibrated and enriched with metadata and AR identifiers. We further use machine-learning methods to evaluate the flare-forecasting skill of RARPs-derived parameters in direct comparison with the 18 SHARPs magnetic field parameters taken from the SHARPs data product headers. In this assessment we intentionally adopt simple, interpretable logistic regression classifiers as a common baseline for both SHARPs and RARPs predictors, leaving more sophisticated machine-learning architectures for future work.
 
To demonstrate RARPs scientific utility, we present case studies including:
\begin{itemize}
    \item flare forecasting using machine learning approaches to compare radio-derived parameters with SHARPs magnetographic parameters for solar flare forecasting;
    \item evolutionary analyses of active-region spectral and polarization curves across their lifetimes.
\end{itemize}

The main contributions of this work are:
\begin{itemize}
    \item development of the RARPs database, providing analysis-ready multi-frequency radio products for NOAA ARs from the long-term RATAN-600 archive;
    \item extension of \texttt{RATANSunPy} with automated AR extraction, calibration, and metadata handling for high-throughput, reproducible processing;
    \item evaluation of the predictive skill of RARPs-based parameters for flare forecasting using machine learning, in comparison with established SHARPs predictors.
\end{itemize}

The structure of the paper is as follows. In the first section, we briefly describe radio spectrum observations, the RATAN-600 instrument, the \texttt{RATANSunPy} library, and its new AR extraction features. Next, we provide a detailed description of the RARPs database and usage examples. Finally, we demonstrate the database’s application in solar flare forecasting using machine-learning models and AR evolution analysis.

\section{RATAN-600 system layout and RATANSunPy version 2}

\subsection{RATAN-600: Architecture, Instrumentation, and Observational Capabilities}

RATAN-600 features a segmented, annular primary reflector, divided into four autonomously controlled segments positioned at cardinal orientations \citep{bogod2011ratan}. The optical system is complemented by a long, Flat secondary mirror inside the ring \citep{tokhchukova2014computation}. This design enables periscope-like configurations in which different sector groupings can be used concurrently, supporting up to three simultaneous observing programs and highly flexible scheduling \citep{tokhchukova2014computation}. Recent upgrades introduced a new spectroradiometric complex with intelligent radio-frequency interference (RFI) suppression, high-speed statistical processing for real-time analysis, and very fine spectral and temporal sampling over 8000 channels per GHz with temporal resolution up to 8 ms, significantly expanding the range of scientifically accessible phenomena \citep{bogod2023spectroradiometry}.

\begin{figure*}[!hbt]
\centering
\includegraphics[width=\linewidth]{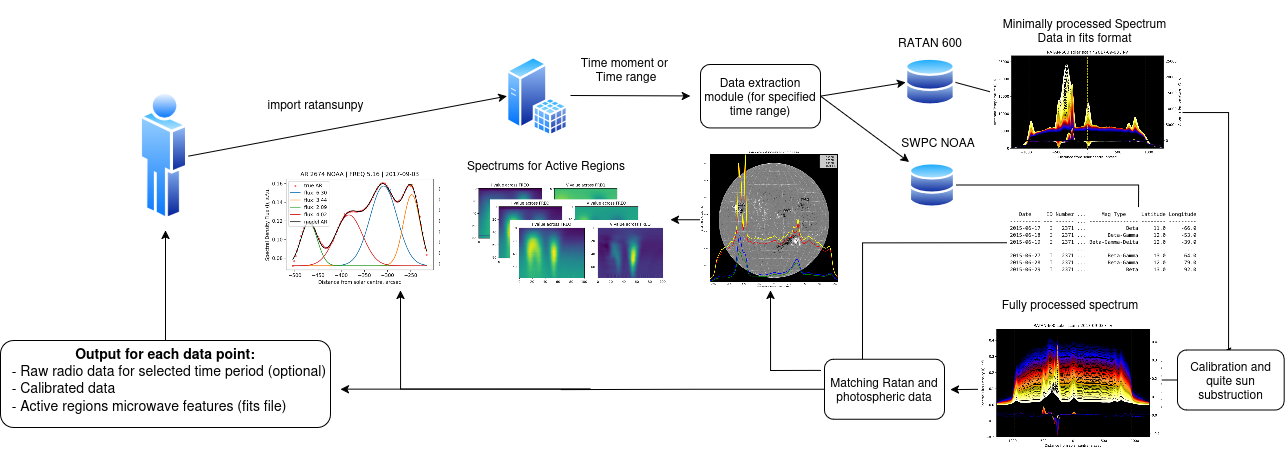}
\caption{Workflow of the \texttt{RATANSunPy} package. The toolkit supports automatic retrieval of raw RATAN-600 solar patrol data, full-disk calibration and preprocessing, extraction of active-region–centered radio spectra, and derivation of per-AR microwave features in standardized FITS format.}
\label{fig:PackageInfo}
\end{figure*}

For solar observations, RATAN-600 employs a three-reflection optical chain. Radiation from the Sun is first intercepted by the internal Flat reflector and directed toward the Southern sector, where the curved panels focus the beam into receiver cabins that can be repositioned along the focal curves \citep{parijskij1993ratan,akhmedov1982measurement}. Coordinated adjustment of the Flat reflector maintains proper coupling to the Southern sector during scans, which is critical for routine solar monitoring. This setup has proven effective across decimeter and centimeter wavelengths. Recent instrumentation upgrades in the decimeter and centimeter bands, coupled with improved RFI mitigation, further enhance spectral and temporal fidelity \citep{ripak2023rfi,bodog2022detection}.

Systematic solar observations began in 1974, with a standardized program established in 1997, creating a multi-decade archive widely used for research \citep{bogod2011ratan,tokhchukova2011ratan}. RATAN-600 observations since 1997 can be accessed at the address of the Prognostic Center of the St. Petersburg branch of the Special Astrophysical Observatory of the Russian Academy of Sciences \footnote{SAO Prognostic Center: \url{http://spbf.sao.ru/prognoz}}. The primary instrument for the centimeter band is an 80-channel spectropolarimeter covering 3–18 GHz, which records right- and left-handed circular polarization (R and L) to derive Stokes parameters:

\begin{equation}
\begin{aligned}
I &= \frac{R + L}{2}, \quad V = \frac{R - L}{2},\\
\text{where:} &\\
I &\text{ is the total intensity},\\
V &\text{ is circular polarization}.
\end{aligned}
\end{equation}

These capabilities are being extended through next-generation spectral complexes across the full operational band of RATAN-600, targeting weak coronal features that require high sensitivity and fine frequency sampling \citep{bogod2025concept}.

\subsection{RATANSunPy version 2}

\texttt{RATANSunPy} is a Python library designed to streamline the acquisition, calibration, and analysis of solar radio observations from the RATAN-600 radio telescope. It adheres to SunPy conventions while accommodating the unique characteristics of RATAN-600's scanning pattern and native data formats. This tool simplifies the complex processing required to work with RATAN-600 data, which is essential for diagnosing solar plasma conditions and predicting solar activity. The main workflow is illustrated in Figure~\ref{fig:PackageInfo}.

\textit{Core Capabilities}

The library's key functionalities are:
\begin{itemize}
    \item \textbf{Calibration:} The calibration pipeline harmonizes metadata, converts data to physical units for Stokes $I$ and $V$, performs baseline removal, and carries out quiet-Sun background subtraction following established practices \citep{borovik1997disser}.
    \item \textbf{AR Detection:} Statistically robust methods are used to detect active region location on full-disk profiles. The pipeline performs artifact rejection and cross-identifies sources with NOAA Solar Region Summary listings to attach AR identifiers.
    \item \textbf{Broadband Microwave Diagnostics:} The library provides tools for frequency-dependent analysis, including estimates of flux density, spectral slopes, and polarization metrics across the 3-18 GHz range, enabling detailed characterization of the microwave properties of ARs \citep{opeikina2015revisiting}.
\end{itemize}

\textit{New Features in Version 2}

Version 2 of \texttt{RATANSunPy} introduces automated detection and extraction of solar ARs. The \texttt{ARHandler} module aligns NOAA heliographic coordinates with scan geometries and extracts user-defined frequency--spatial windows for each region, enabling consistent time--frequency analysis on a per-region basis. The \texttt{ARClient} offers access to a curated repository of pre-extracted active-region products, facilitating statistical studies and large-scale analyses. Quality control procedures include outlier rejection, adaptive masking of bright artifacts. The resulting outputs comprise calibrated scan objects and tabulated region and source parameters, ready for downstream analysis. 

\subsection{Automatic detection and extraction of ARs}

\begin{figure*}[ht]
\centering
\includegraphics[width=1\linewidth]{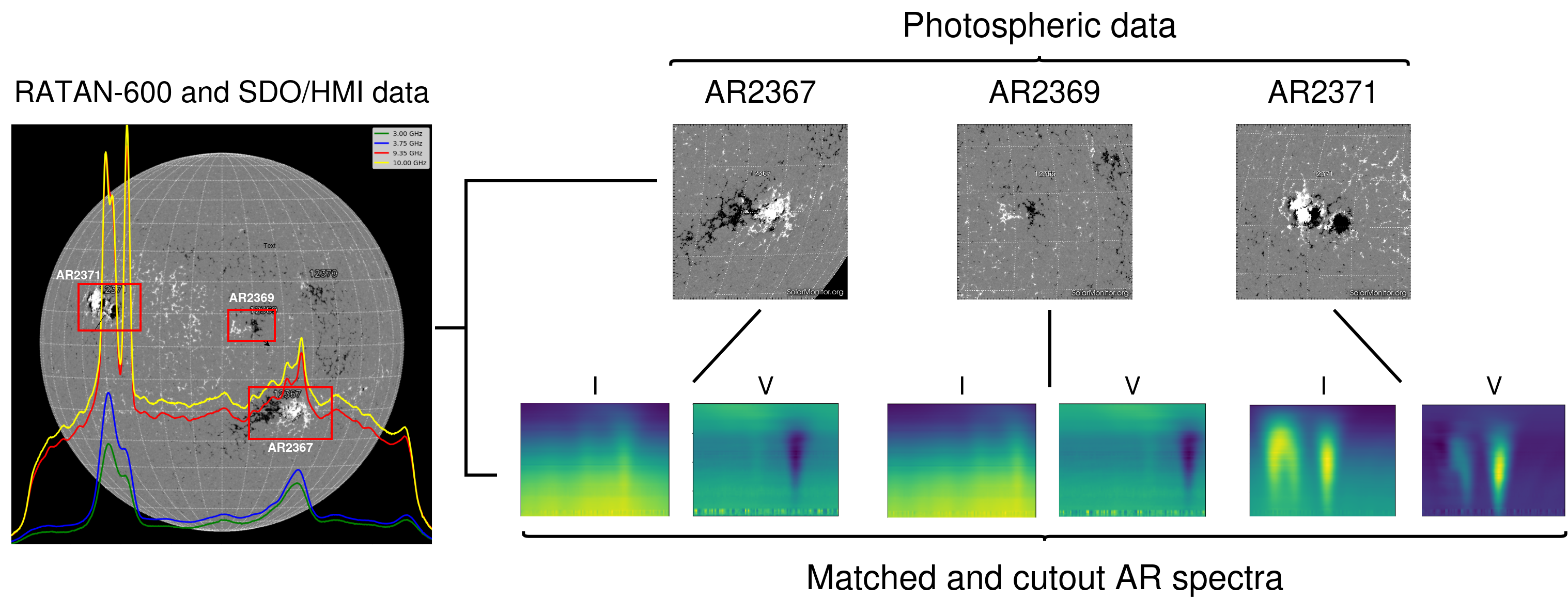}
\caption{Schema of AR spectra extraction. The left panel shows how full-disk photospheric data from SDO/HMI are spatially aligned with selected RATAN-600 full-disk radio scans at 3.00 GHz (green), 3.75 GHz (blue), 9.35 GHz (red), and 10.00 GHz (yellow). The right panel illustrates how photospheric active-region patches are matched with their corresponding cutouts extracted from the RATAN-600 full-disk radio spectrum. Image was taken from \href{https://sdo.gsfc.nasa.gov/assets/img/latest/}{the SDO observatory site}}
\label{fig:SDO-to-Ratan}
\end{figure*}

\texttt{RATANSunPy} v2 brings sophisticated AR extraction and localization capabilities through the \texttt{ARHandler} class. This new functionality addresses the critical gap between raw full-disk radio observations and analysis-ready AR-specific data products. Figure~\ref{fig:SDO-to-Ratan} schematically summarizes the AR spectra extraction procedure, from the spatial alignment of SDO/HMI photospheric data with RATAN-600 radio scans to the matching of active-region patches with their corresponding radio cutouts.

The extraction process begins with the identification of active regions (ARs) using the NOAA Solar Region Summary (SRS) reports, which provide heliographic coordinates and magnetic classifications. In all extracted AR spectra, the two data axes correspond to the observing frequency and the spatial offset from the AR center along the one-dimensional RATAN-600 scan. The spatial axis consists of uniformly spaced scan pixels with an angular step and a reference position given by the FITS keywords \texttt{CDELT1} (arcsec per pixel) and \texttt{CRPIX1} (reference pixel).

For each patch we extract a fixed--size window centered on the active region so that a specific pixel index (for example, $i=50$) marks the AR centre in the scan frame. Using the information stored in the FITS header (the keywords \texttt{CDELT1} and \texttt{CRPIX1}, together with \texttt{SOLAR\_R}, \texttt{SOLAR\_B} and \texttt{ANGLE}), each spatial pixel index can be converted to a helioprojective coordinate (in arcseconds) and subsequently to heliographic latitude and longitude. The system then performs the following key steps.

\begin{itemize}
    \item \textbf{Coordinate Alignment and Localization:}  
        The \texttt{ARHandler} aligns heliographic coordinates from NOAA SRS reports with the RATAN-600 scan axis, enabling consistent spatial registration between reported AR positions and the radio scan data.

    \item \textbf{Spectral Window Extraction:}  
        For each identified AR, the system extracts a configurable spatial window (default: ±50 pixels) around the AR center from the full-disk radio scan. This produces a 2D spectrum with dimensions [frequency × spatial] that captures the radio emission characteristics specific to that AR.
    
    \item \textbf{Automated Quality Control:}  
        The extraction process includes outlier detection and replacement algorithms that identify and correct anomalous data points based on statistical thresholds. Outliers exceeding the 99th percentile are replaced by a one-dimensional local linear interpolation along the scan (spatial) direction, implemented as the arithmetic mean of the two neighboring pixels at the same frequency channel.
    
    \item \textbf{AR Mask Generation:}  
        A binary mask is automatically generated to identify the most radio-bright regions within each AR patch. This mask uses adaptive thresholding followed by morphological hole-filling to create clean regions of interest.

    \item \textbf{Statistical Characterization:}  
        For each AR patch, the system computes comprehensive statistical descriptors including mean, standard deviation, minimum, maximum, and integrated flux values across all frequency channels. These statistics are computed both globally and within the AR mask region.

    \item \textbf{Metadata Integration:}  
        Each extracted AR includes rich metadata from multiple sources: observational parameters (date, time, azimuth, solar coordinates), magnetic field classifications (McIntosh class, magnetic type) from SRS reports, and nearby AR information for context.
\end{itemize}

This automated pipeline transforms raw RATAN-600 scans into standardized, analysis-ready AR data products, dramatically reducing the technical barrier for researchers to access and analyze radio signatures of solar activity.

\subsection{RATANSunPy v2 and AR Extraction Summary}

\texttt{RATANSunPy} v2 automates the transformation of raw RATAN-600 full-disk scans into  AR-specific radio data products through robust calibration, coordinate-aligned extraction, and statistical characterization. The \texttt{ARHandler} module addresses key challenges in handling scan geometries and data quality, enabling reproducible, high-throughput analysis of multi-frequency spectra and polarization for individual active regions or large-scale studies. This functionality lowers barriers to accessing coronal radio diagnostics, facilitating integration with photospheric datasets like SHARPs for advanced solar activity research.

\section{Database description, usage examples}

\subsection{Database Overview and Coverage}
The RARPs database represents the first comprehensive, standardized collection of solar AR radio signatures derived from RATAN-600 observations. The database currently spans from 2009-01-10 to 2025-08-31.

Each entry in RARPs corresponds to a single AR observed during one RATAN-600 scan, providing multi-frequency radio spectra from 3-18 GHz. During a typical observing day, the RATAN-600 solar patrol acquires a sequence of full-disk scans at multiple azimuths spanning a regular grid (e.g., from $+30^\circ$ to $-30^\circ$), each of which can contain several NOAA active regions. For a given active region, the RARPs pipeline extracts one spectrum per azimuth position per day — each azimuth step is used once, without repetition. This results in multiple observations of each AR per day, one for each position in the azimuth scan grid. The temporal structure is thus inherently azimuth-driven: the natural sampling cadence is uniform steps in azimuth (typically $2^\circ$ per spectrum), rather than a uniform temporal interval. Occasional gaps occur due to weather or technical downtime, but when observations are acquired, they follow this regular azimuth grid pattern. The database includes observations of ARs ranging from simple bipolar configurations to complex multi-polar systems, covering the full range of magnetic activity levels from quiet emergence to major flare-productive regions.

The temporal coverage of RARPs provides opportunities for studying AR evolution across multiple timescales, from short-term (hours to days) spectral variations to long-term (months to years) statistical trends. This comprehensive coverage enables both individual case studies of specific ARs and large-scale statistical analyses of AR populations.

\subsection{AR FITS File Structure and Content}

Each AR observation is stored as a self-contained FITS file following astronomical data standards. The FITS structure consists of multiple Header Data Units (HDUs) designed for efficient data access and analysis:

\begin{itemize}
    \item \textbf{Primary HDU (Spectrum Data):}  
    Contains the core 3D array with dimensions \([3 \times \text{frequency} \times \text{spatial}]\), where the three layers represent:
    \begin{itemize}
        \item \textbf{Layer 0:} Intensity (Stokes I, K) measurements
        \item \textbf{Layer 1:} Circular polarization (Stokes V, K) measurements
        \item \textbf{Layer 2:} Binary mask (boolean type) identifying significant emission regions
    \end{itemize}
    Here, the \textit{frequency} axis indexes the receiver channels between 3–18~GHz, while the \textit{spatial} axis indexes the one-dimensional sequence of scan pixels that form the extracted active-region patch. Both Stokes I and V are measured in brightness temperature units (kelvin, K).
    
    \item \textbf{Statistical HDUs:}  
    Additional HDUs contain pre-computed statistical arrays for efficient analysis:
    \begin{itemize}
        \item \textbf{MEAN:} Frequency-dependent mean values
        \item \textbf{STD:} Standard deviation arrays
        \item \textbf{MIN:} Minimum values per frequency channel
        \item \textbf{MAX:} Maximum values per frequency channel
        \item \textbf{SUM:} Integrated flux measurements
        \item \textbf{FREQ:} Frequency array in GHz
    \end{itemize}

    \item \textbf{Header Metadata:} Each FITS file header contains comprehensive metadata essential for scientific analysis (see Table~\ref{tab:fits_header}).
\end{itemize}

\renewcommand{\arraystretch}{1.3}
\begin{table*}[ht]
\centering
\caption{RARPs FITS Header Parameters for NOAA AR~12371}
\label{tab:fits_header}
\scriptsize
\begin{tabularx}{\textwidth}{@{}lXl@{}}
\toprule
\textbf{Header Key} & \textbf{Description} & \textbf{Example Value} \\
\midrule
AR\_NUM   & NOAA Active Region Number                & 12371 \\
LATITUDE  & Heliographic Latitude (arcsec)           & -478.8954 \\
DATE-OBS  & Observation Date                         & 2015/06/19 \\
TIME-OBS  & Observation Time (UT)                    & 09:14:53.440 \\
AZIMUTH   & Telescope Azimuth (degrees)              & 0.0 \\
CDELT1    & Spatial Resolution (arcsec/pixel)        & 2.8914 \\
CRPIX1     & Reference Pixel                          & 527.2591 \\
SOLAR\_R  & Solar Radius (pixels)                    & 944.47 \\
SOLAR\_B  & Solar P-angle (degrees)                  & 1.5 \\
SOL\_DEC  & Solar Declination (degrees)              & 23.4150 \\
ANGLE     & Position Angle (degrees)                 & -8.1 \\
MAG\_TYPE & Magnetic Classification                  & Beta-Gamma-Delta \\
MCINTOSH  & McIntosh Classification                  & Ekc \\
LAT\_INTR & Overlap in latitude with neighboring ARs & NONE \\
\bottomrule
\end{tabularx}
\end{table*}

This standardized format ensures compatibility with astronomical analysis software while providing all necessary information for coordinate transformations, temporal analysis, and multi-wavelength studies.

\subsection{Database Access and Usage Examples}
The RARPs database is designed for easy programmatic access through the extended \texttt{RATANSunPy} package. The \texttt{ARClient} class provides a high-level interface for querying and downloading AR data with filtering capabilities.

\textit{Basic Data Acquisition}

The following example shows how to collect all data for the specified timerange:

\begin{lstlisting}[language=Python, breaklines=true, basicstyle=\ttfamily\scriptsize]
from ratansunpy.client import ARClient
from ratansunpy.time import TimeRange

# Initialize client
ar_client = ARClient()
# Define time range
timerange = TimeRange('2015-01-01', '2015-12-31')
# Specify active regions numbers
ar_numbers = ["12371", "12369", "12367"]
# Download data
ar_local_paths = ar_client.download_data(
    timerange=timerange, 
    ar_nums=ar_numbers, 
    save_to="./ar_data")
\end{lstlisting}

\textit{Advanced Filtering}

The client provides additional functionality to filter data by AR number and azimuth, with a local caching mechanism:

\begin{lstlisting}[language=Python, breaklines=true, basicstyle=\ttfamily\scriptsize]
# Filter by time range 
timerange = TimeRange('2015-06-01', '2015-08-31')
# Filter by active regions numbers
ar_numbers = ["12371", "12370"]
# Filter by azimuth range
azimuths = [30, 20, 10, 0, -10, -20, -30]
ar_local_paths = ar_client.download_data(timerange=timerange,
                                         ar_nums=ar_numbers,
                                         azimuths=azimuths,
                                         cache=True,
                                         save_to="./ar_data")
\end{lstlisting}

\textit{Data Processing and Analysis} 

Once downloaded, FITS files can be processed using standard astronomical libraries:

\begin{lstlisting}[language=Python, breaklines=true, basicstyle=\ttfamily\scriptsize]
from astropy.io import fits

# Read AR data 
hdul = fits.open(ar_local_paths[0])
# Access spectrum data
intensity = hdul[0].data[0]       # I channel
polarization = hdul[0].data[1]    # V channel
mask = hdul[0].data[2]            # AR mask

# Access pre-computed statistics
mean_spectrum = hdul['MEAN'].data
frequencies = hdul['FREQ'].data

# Extract metadata
ar_number = hdul[0].header['AR_NUM']
observation_time = hdul[0].header['TIME-OBS']
magnetic_type = hdul[0].header['MAG_TYPE']
\end{lstlisting}

\subsection{Data Re-extraction and Customization}

\texttt{RATANSunPy} also provides tools for flexible data extraction and fine-grained control of the processing parameters.  
This functionality is particularly useful when the default AR extraction does not fully capture the structures of interest, or when researchers want to experiment with different spatial resolutions and sensitivity thresholds.  

\textit{Custom Window Sizes}

AR data can be extracted with user-defined window sizes, controlling the number of spatial pixels around the AR latitude:

\begin{lstlisting}[language=Python, breaklines=true, basicstyle=\ttfamily\scriptsize]
from ratansunpy.core.ar_handling import ARHandler
from ratansunpy.client import RATANClient

# Initialize RATANClient 
ratan_client = RATANClient()

# Extract and process raw RATAN data for the whole sun disk
urls = ratan_client.acquire_data(timerange)
_, calibrated_data = ratan_client.process_fits_data(urls[0])

# Form SRS table for matching latitude and active regions
srs_table = ratan_client.form_srstable_with_time_shift(calibrated_data)

# Initialize handler with custom window size
handler = ARHandler(calibrated_data)

# Extract custom active region data    
ar_data = handler.extract_ar_data_with_window(
    latitude=-478.9,
    window_size=100)
\end{lstlisting}

\textit{Alternative Processing Parameters}  

Furthermore, users can customize the AR extraction process by adjusting parameters such as \texttt{window\_size}, \texttt{ar\_number} for explicitly targeting a specific region, and \texttt{threshold\_multiplier}, which controls sensitivity to intensity variations.

\begin{lstlisting}[language=Python, breaklines=true, basicstyle=\ttfamily\scriptsize]
# Custom processing with different parameters
ar_hdul, filename = handler.process_one_region(
    latitude=-478.9,
    ar_number="12371",
    window_size=75,             # default: 50
    threshold_multiplier=3.0)   # default: 2.5
\end{lstlisting}

This flexible system ensures that RARPs can accommodate diverse research requirements while maintaining a standardized database structure.

Comprehensive usage examples can be found in the example notebooks available on the project's \texttt{github} repository \footnote{\protect\url{https://github.com/SpbfSAO/RATANSunPy}} and within the official documentation \footnote{\protect\url{https://spbfsao.github.io/RATANSunPy/}}. 

\section{Descriptive analysis}

\subsection{Machine Learning Approach and Data Preparation}
To demonstrate the scientific utility of the RARPs database for space weather applications, we conducted a comprehensive comparison of radio-based versus magnetic field-based solar flare forecasting. This analysis utilized data from 2019-01-02 to 2025-04-15, covering the ascending phase of solar cycle 25, and employed machine learning techniques to extract features from both radio spectra and traditional magnetic field parameters. In total, the forecasting dataset comprises tens of thousands of active-region observations with matched SHARPs parameters, RARPs radio spectra, and GOES flare labels; a quantitative overview of the final database and the specific subsets used for training and testing the models is provided in Table~\ref{tab:rarp_db} and Table~\ref{tab:rarp_ml} below. The complete forecasting pipeline—from data sources and synchronization to evaluation—is summarized in Table~\ref{tab:forecast_pipeline_summary}.

\begin{table}[h]
\centering
\caption{RARPs database: coverage statistics.}
\label{tab:rarp_db}
\scriptsize
\begin{tabular}{p{0.6\columnwidth} >{\centering\arraybackslash}p{0.3\columnwidth}}
\toprule
\textbf{Quantity} & \textbf{Value} \\
\midrule
Total AR samples (patches)  & 162,832 \\
Unique AR (NOAA numbers)    & 2,884 \\
Total time coverage         & 2009-01-10 to 2025-08-31 \\
RATAN-600 scans             & 38,466 \\
\bottomrule
\end{tabular}
\end{table}

\begin{table}[h]
\centering
\caption{ML dataset composition and flare-positive sample counts.}
\label{tab:rarp_ml}
\scriptsize
\begin{tabular}{p{0.6\columnwidth} >{\centering\arraybackslash}p{0.3\columnwidth}}
\toprule
\textbf{Quantity} & \textbf{Value} \\
\midrule
Total samples               & 53,197 \\
Training samples            & 46,948 \\
Test samples                & 6,249 \\
C-class positive (0--24 h)  & 9,499 \\
M$\geq$ positive (0--24 h)  & 3,326 \\
C-class positive (24--48 h) & 8,540 \\
M$\geq$ positive (24--48 h) & 2,899 \\
\bottomrule
\end{tabular}
\end{table}

We synchronized RARPs observations with GOES flare catalogs and HMI SHARPs magnetic field measurements. For each RARPs spectrum, identified by its observation time and NOAA active-region number, we retrieved the corresponding SHARPs magnetic parameters by mapping the NOAA number to its HMI Active Region Patch Number (HARP)  identifier. Given the high 12-minute cadence of SDO/HMI relative to RATAN-600 observations, we assigned the strictly nearest available SHARPs snapshot to each radio spectrum, ensuring precise temporal alignment. Observations without a corresponding SHARPs record were excluded from the comparative analysis.
Flare labels were assigned by cross-matching each RARPs entry with the GOES flare list based on the NOAA active-region number and observation time. For each radio spectrum, we identified all flares occurring within two subsequent prediction windows (0–24h and 24–48h) and utilized their classifications (e.g. C1.0, M2.5) to define binary targets for C1.0–C9.9 and M1.0+ (including X-class) flares.

\begin{figure}[!hbt]
\centering
\includegraphics[width=1\linewidth]{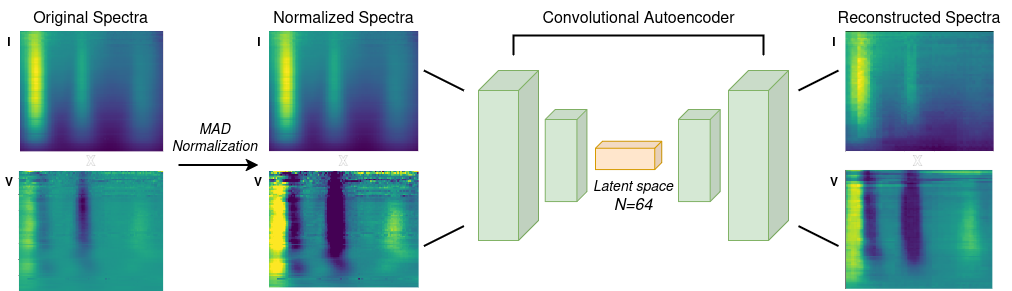}
\caption{Convolutional autoencoder schema with examples of reconstructed data.}
\label{fig:ConvAE}
\end{figure}

A key challenge in utilizing radio data for machine-learning applications is the high dimensionality and complex structure of the 2D spectral images: each RARPs patch is a two-channel array (Stokes $I$ and $V$) of size 75×101, corresponding to 15,150 spatial–spectral features per active region if treated as a flat vector. To address this, we applied a convolutional autoencoder architecture specifically designed for RATAN-600 AR spectra (see Figure~\ref{fig:ConvAE}). This represents, to our knowledge, the first use of deep autoencoder-based feature extraction on a large, standardized database of solar radio spectra for active-region flare forecasting, made possible by the volume and uniformity of the newly constructed RARPs archive. The autoencoder compresses each 2D spectrum into a 64-dimensional latent embedding that serves as a compact, data-driven representation of the underlying coronal radio emission.

The autoencoder architecture consists of:

\begin{itemize}
\item Encoder: Three convolutional layers (32, 64, 128 filters) with batch normalization, ReLU activation, and max-pooling
\item Bottleneck: Dense layer compressing to 64-dimensional latent space
\item Decoder: Three transposed convolutional layers reconstructing original spectra
\end{itemize}

Before training the autoencoder, the RARPs radio data were preprocessed by cropping each calibrated active-region map to a fixed spatial size, replacing rare missing values in Stokes $I$ and $V$ with zeros, no additional selection or correction based on telescope azimuth was applied in this baseline preprocessing. Additionally, to mitigate heavy-tailed artifacts, we implemented Median Absolute Deviation (MAD) normalization, which is robust to outliers:

\begin{equation}
\begin{aligned}
x^{\mathrm{norm}}_i &= \frac{x_i - \operatorname{median}(x)}{\operatorname{MAD}(x)},\\
\text{where}\quad \operatorname{MAD}(x) &= \operatorname{median}\bigl(|x_i - \operatorname{median}(x)|\bigr)
\end{aligned}
\end{equation}

After clipping outliers at ±6 MAD, data were normalized using global statistics.

\begin{figure}[!hbt]
\centering
\includegraphics[width=1\linewidth]{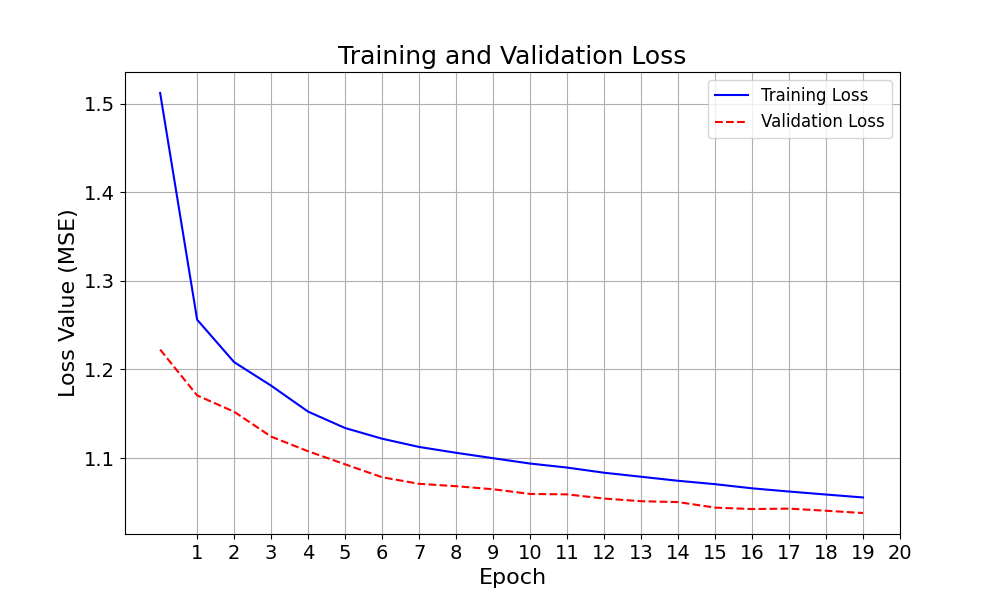}
\caption{Training and validation loss curves of the convolutional autoencoder over 20 epochs.}
\label{fig:AE_losses}
\end{figure}

Training and validation losses decreased steadily (Fig.~\ref{fig:AE_losses}), with training loss improving from 1.51 to 1.06 and validation loss from 1.22 to 1.04 over 20 epochs, indicating stable learning without overfitting. The trained autoencoder preserved key structures while reducing noise, as reflected in the reconstruction metrics:

\begin{itemize}
\item Intensity (I) channel: PSNR = 29.33 dB, SSIM = 0.803
\item Polarization (V) channel: PSNR = 21.28 dB, SSIM = 0.611
\end{itemize}

The autoencoder effectively reconstructs the intensity (I) channel, capturing the main spatial structures and temporal patterns of the ARs. Minor distortions occur primarily in low-intensity regions, but overall structural fidelity remains high. For the polarization (V) channel, reconstructions are noisier and less precise, reflecting the inherently higher variance and lower signal-to-noise ratio of circular polarization measurements. Visual inspection shows that fine-scale features in the V channel are smoothed or partially lost, which reduces SSIM and PSNR relative to the intensity channel.

\subsection{Forecasting Metrics}

\begin{figure}[h]
    \centering
    \begin{subfigure}{0.48\linewidth}
        \includegraphics[width=\linewidth]{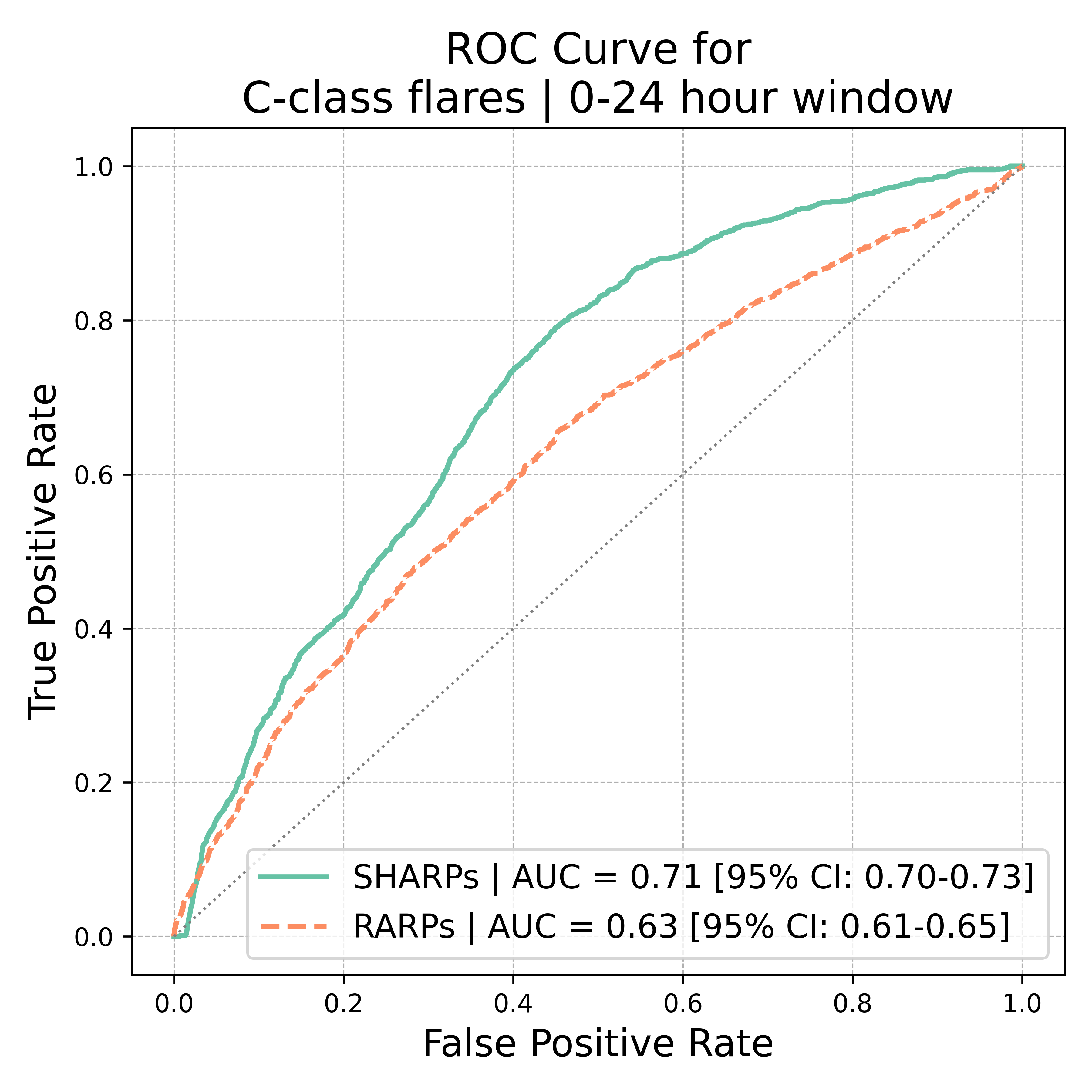}
        \caption{}
        \label{fig:subfig1}
    \end{subfigure}
    \hfill
    \begin{subfigure}{0.48\linewidth}
        \includegraphics[width=\linewidth]{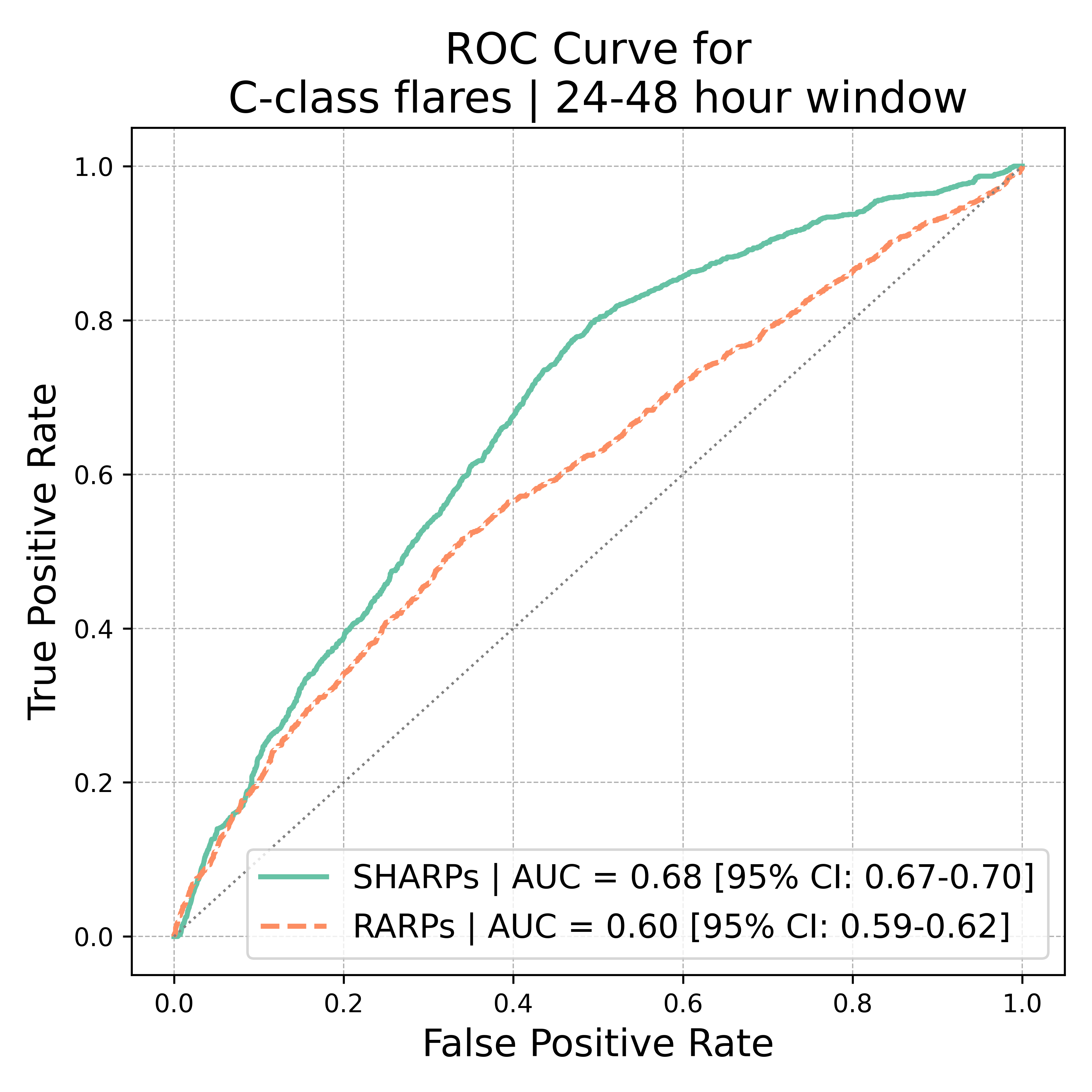}
        \caption{}
        \label{fig:subfig2}
    \end{subfigure}

    \begin{subfigure}{0.48\linewidth}
        \includegraphics[width=\linewidth]{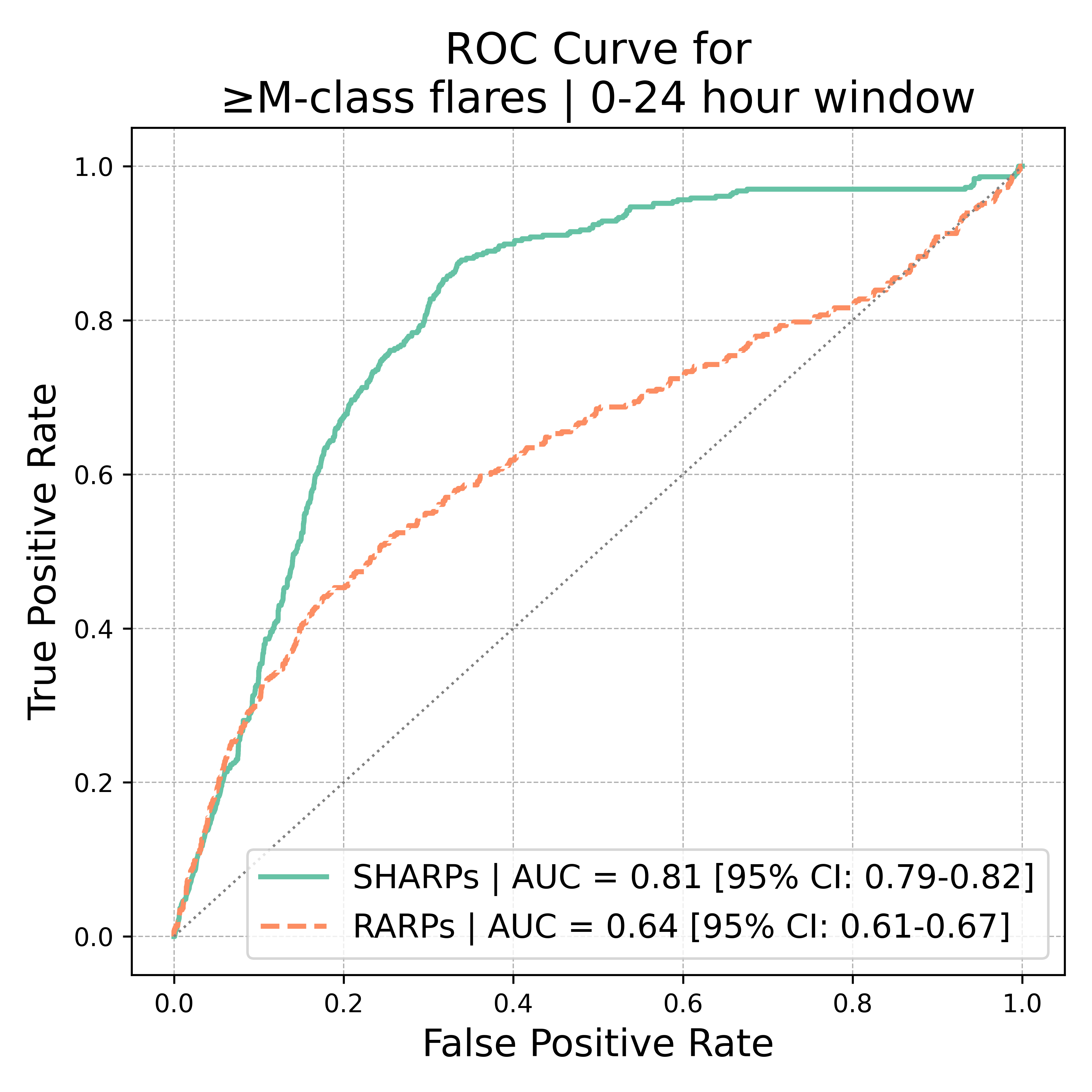}
        \caption{}
        \label{fig:subfig3}
    \end{subfigure}
    \hfill
    \begin{subfigure}{0.48\linewidth}
        \includegraphics[width=\linewidth]{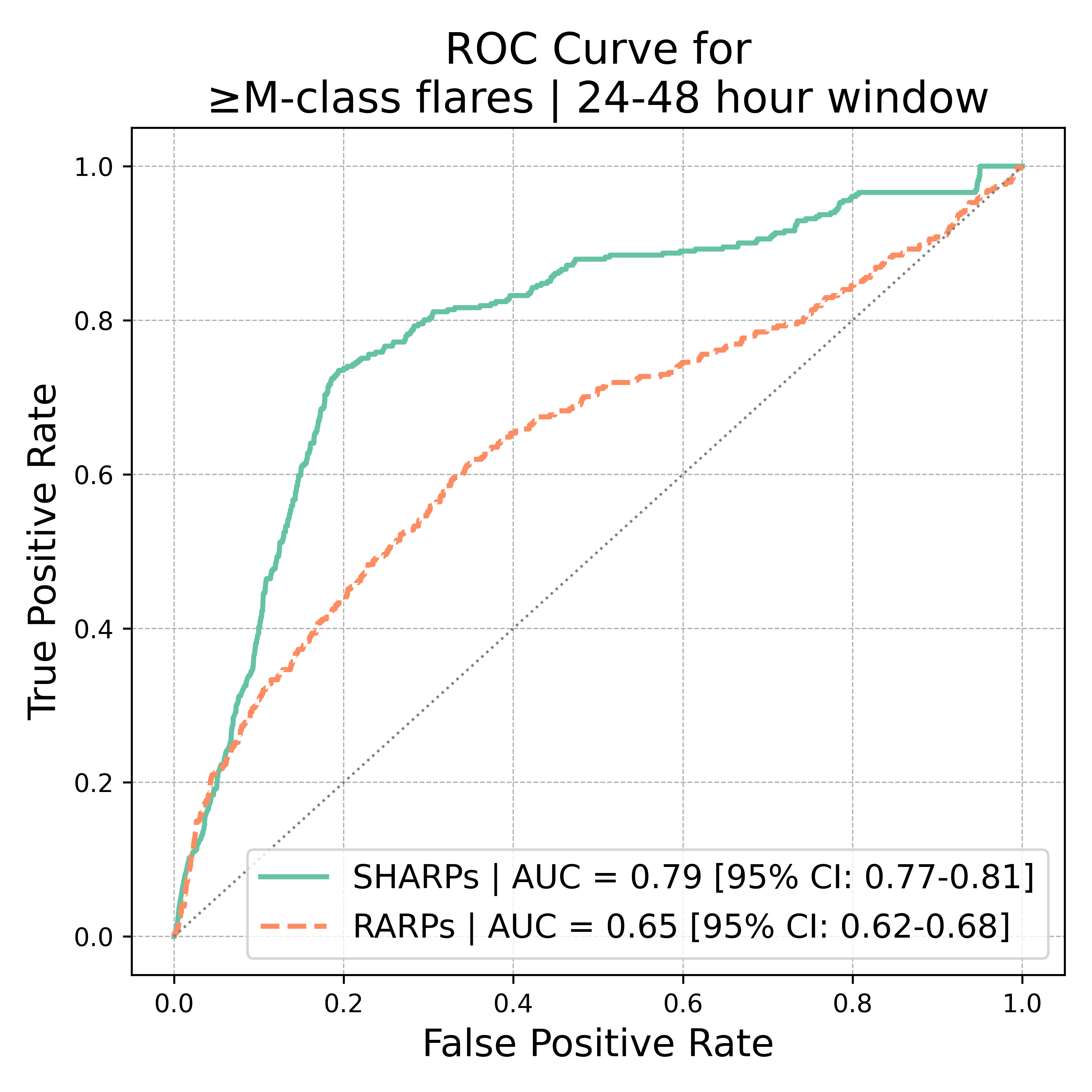}
        \caption{}
        \label{fig:subfig4}
    \end{subfigure}
    
    \caption{Comparison of ROC-AUC scores for Logistic Regression models on SHARPs and RARPs datasets across flare types and prediction windows.}
    \label{fig:roc_overall}
\end{figure}

\begin{figure}[!hbt]
    \centering

    \begin{subfigure}{0.48\linewidth}
        \includegraphics[width=\linewidth]{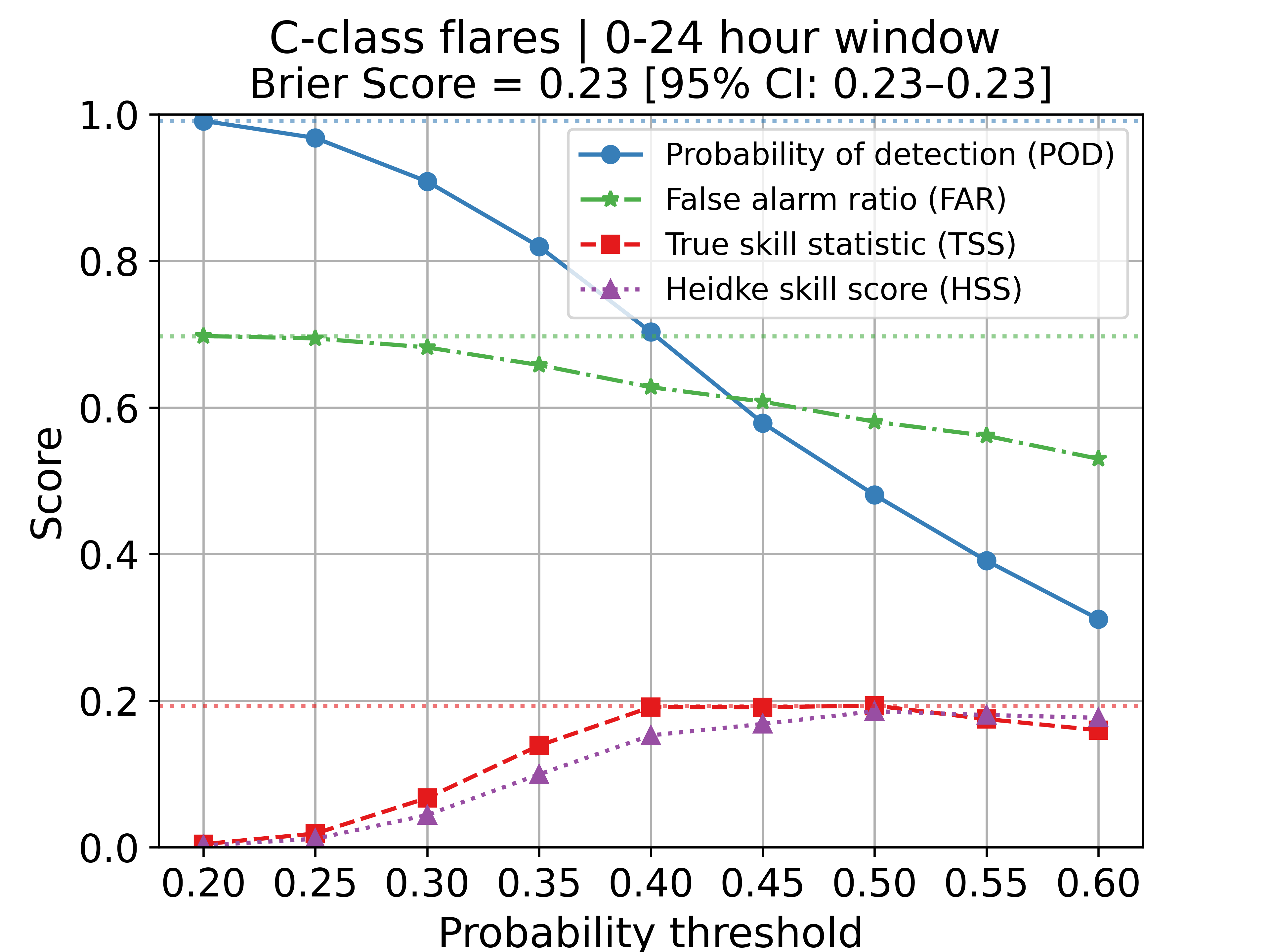}
        \caption{}
        \label{fig:subfig1ar}
    \end{subfigure}
    \hfill
    \begin{subfigure}{0.48\linewidth}
        \includegraphics[width=\linewidth]{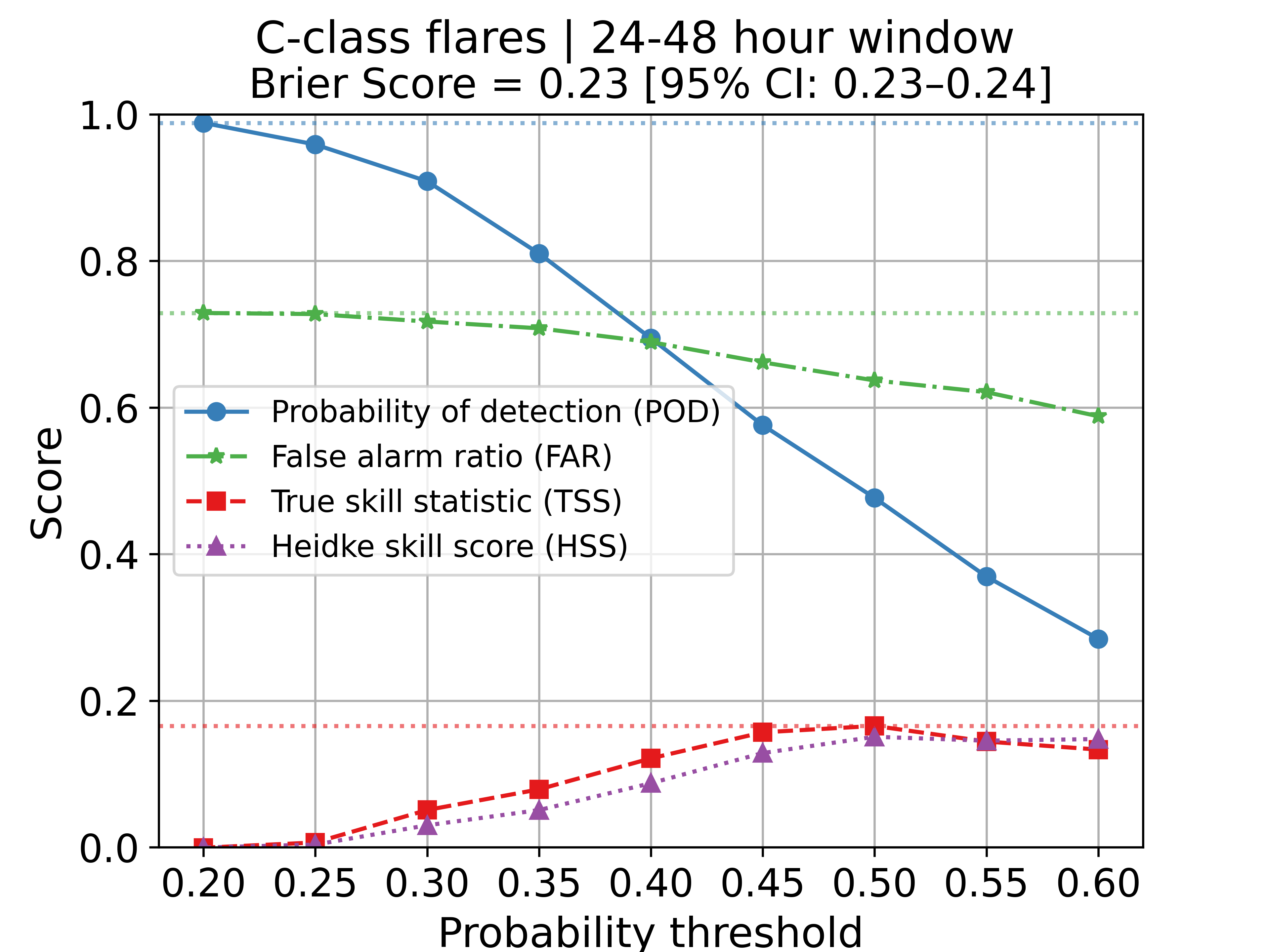}
        \caption{}
        \label{fig:subfig2br}
    \end{subfigure}

    \begin{subfigure}{0.48\linewidth}
        \includegraphics[width=\linewidth]{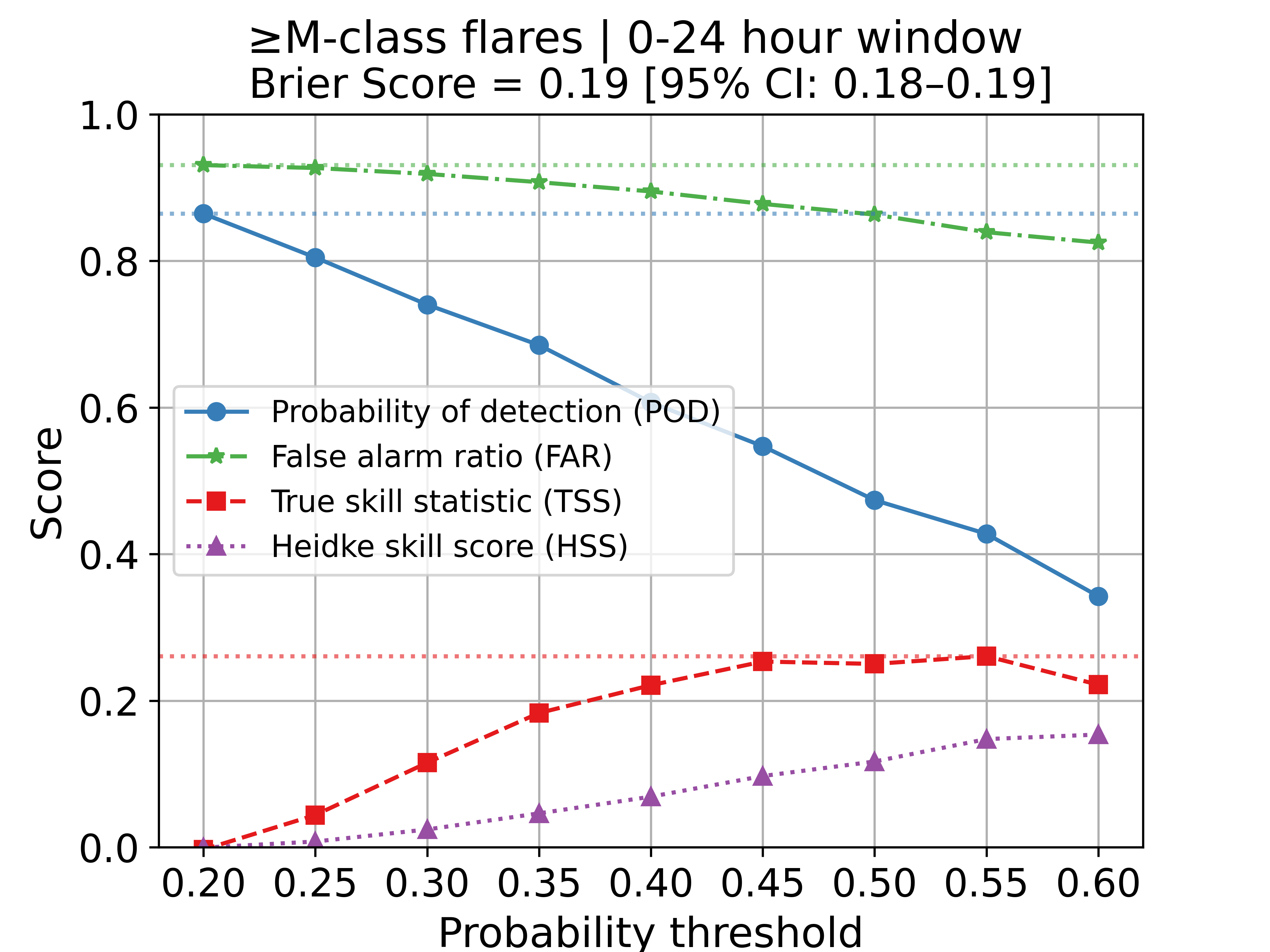}
        \caption{}
        \label{fig:subfig3cr}
    \end{subfigure}
    \hfill
    \begin{subfigure}{0.48\linewidth}
        \includegraphics[width=\linewidth]{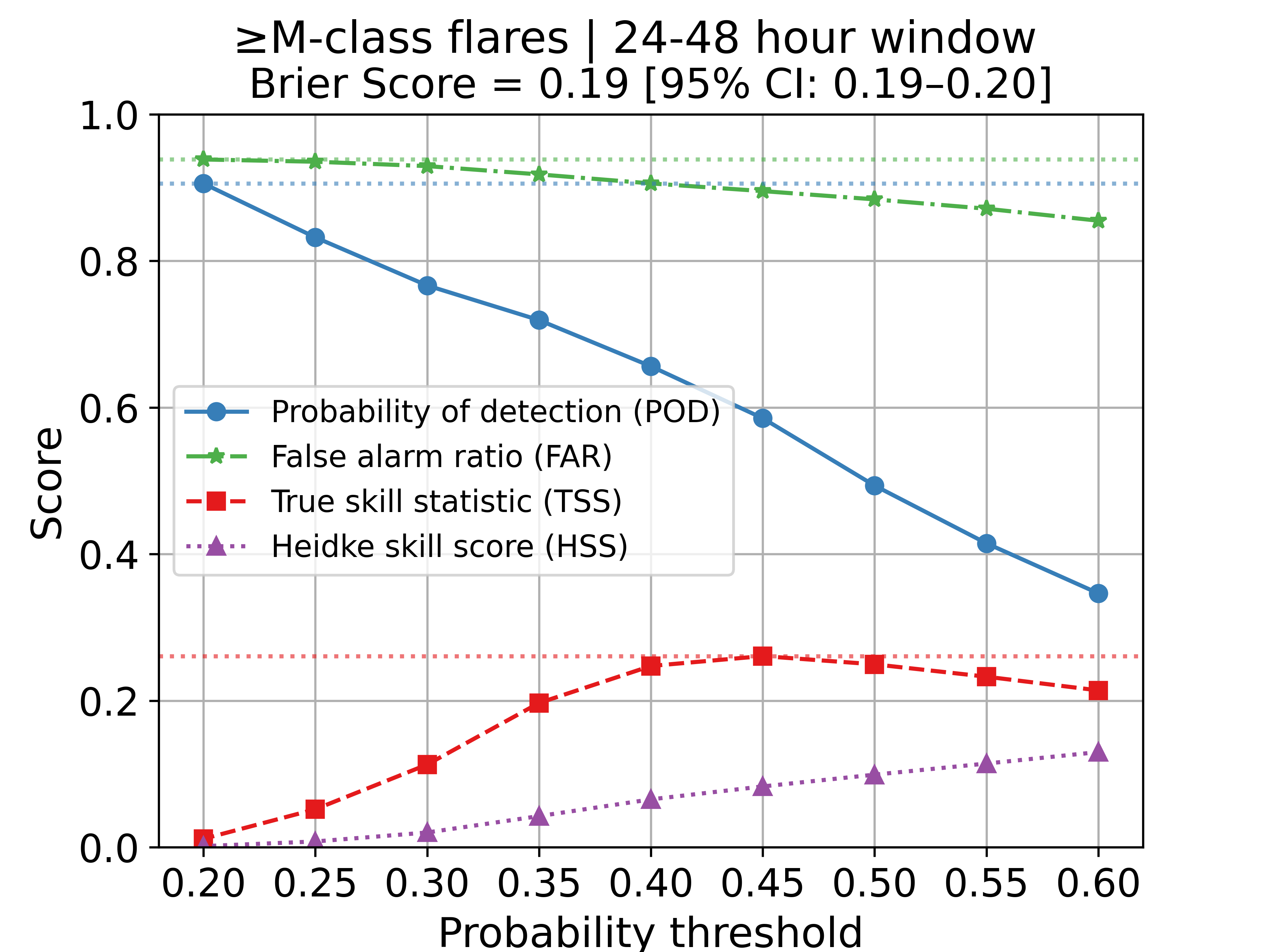}
        \caption{}
        \label{fig:subfig4dr}
    \end{subfigure}
    
    \caption{Threshold-dependent evaluation metrics for the \textbf{RARPs model} baseline logistic regression model predicting solar flare activity.}
    \label{fig:forecasting_ratan}
\end{figure}

\begin{figure}[!hbt]
    \centering

    \begin{subfigure}{0.48\linewidth}
        \includegraphics[width=\linewidth]{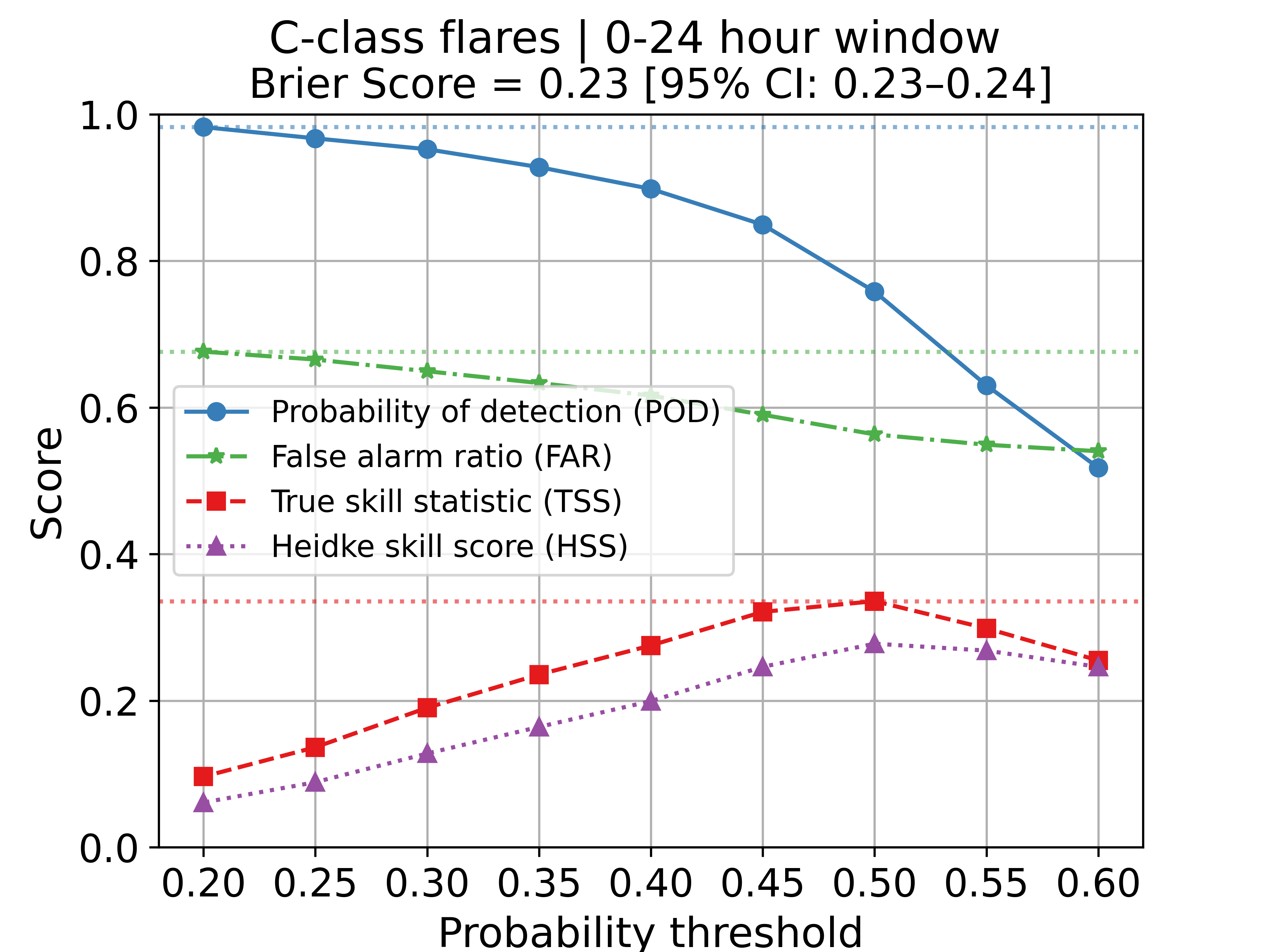}
        \caption{}
        \label{fig:subfig1as}
    \end{subfigure}
    \hfill
    \begin{subfigure}{0.48\linewidth}
        \includegraphics[width=\linewidth]{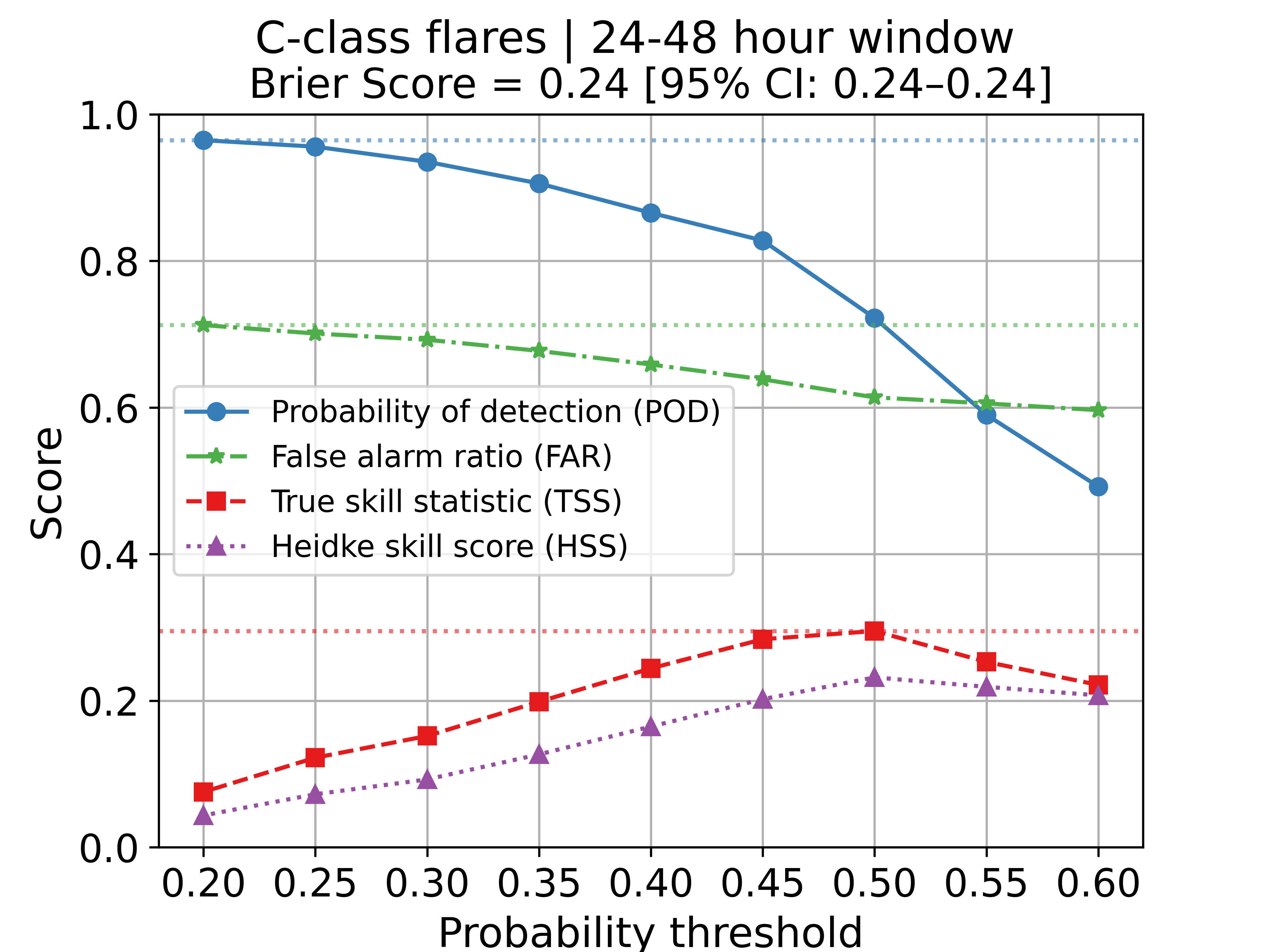}
        \caption{}
        \label{fig:subfig2bs}
    \end{subfigure}

    \begin{subfigure}{0.48\linewidth}
        \includegraphics[width=\linewidth]{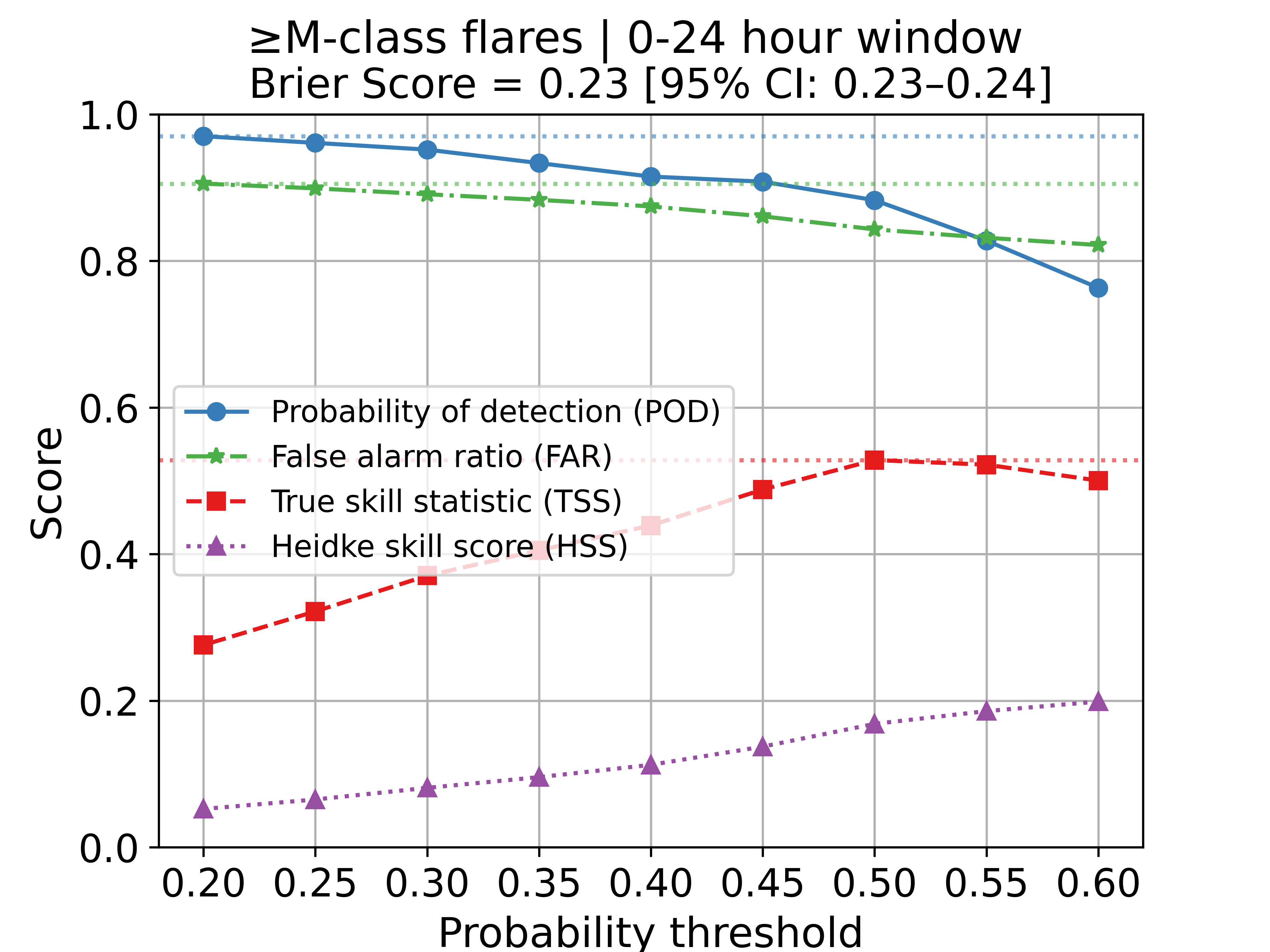}
        \caption{}
        \label{fig:subfig3cs}
    \end{subfigure}
    \hfill
    \begin{subfigure}{0.48\linewidth}
        \includegraphics[width=\linewidth]{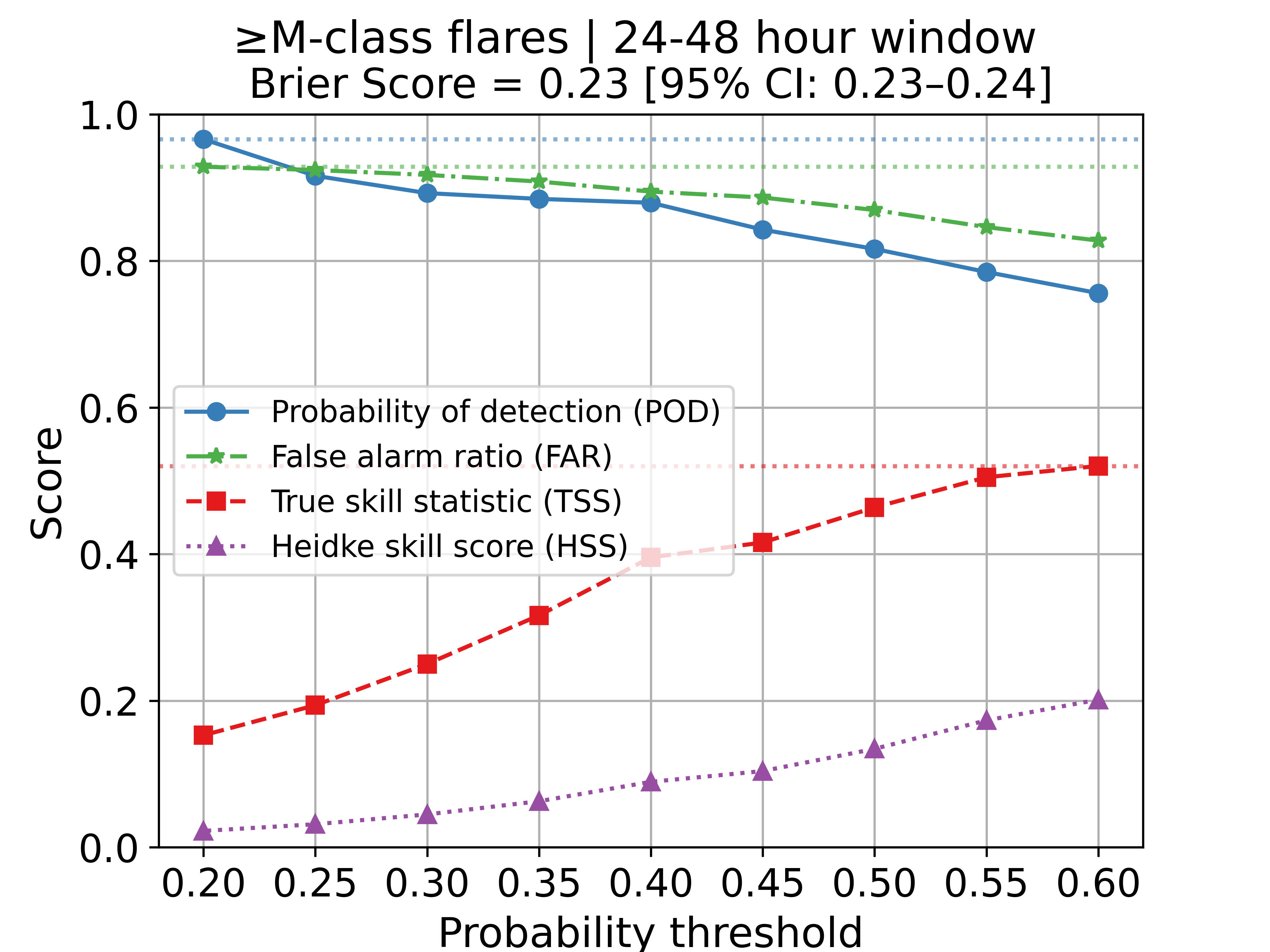}
        \caption{}
        \label{fig:subfig4ds}
    \end{subfigure}
    
    \caption{Threshold-dependent evaluation metrics for the \textbf{SHARPs model} baseline logistic regression model predicting solar flare activity.}
    \label{fig:forecasting_sharp}
\end{figure}

To evaluate solar flare forecasts in a consistent way, we adopt a set of verification metrics standard in the flare-forecasting literature, following the framework established by \cite{bloomfield2012toward}. In our implementation, non-parametric bootstrap resampling was used to estimate uncertainty for the probabilistic metrics ROC-AUC and Brier score, yielding 95\% confidence intervals from the empirical bootstrap distributions. The resulting metrics and their confidence intervals (for ROC-AUC and Brier score) are presented in Figures~\ref{fig:roc_overall}–\ref{fig:forecasting_sharp} and summarized numerically in Table~\ref{tab:summary_all}.

The core metrics are described in detail below, each with its corresponding formula.

\begin{itemize}
    \item \textbf{ROC-AUC}: Overall discriminative ability of the classifier, measured as the area under the Receiver Operating Characteristic curve, which is a probability curve that plots the True Positive Rate (TPR) against the False Positive Rate (FPR) at different thresholds:
        \begin{equation}
        \begin{aligned}
        \text{where}\quad \text{TPR} &= \frac{TP}{TP+FN}, \quad 
        \text{FPR} = \frac{FP}{FP+TN}.
        \end{aligned}
        \end{equation}
    
    \item \textbf{Probability of Detection (POD)}: Fraction of correctly predicted events:
        \begin{equation}
        POD = \frac{TP}{TP + FN}.
        \end{equation}
    
    \item \textbf{False Alarm Ratio (FAR)}: Fraction of predicted events that did not occur:
        \begin{equation}
        FAR = \frac{FP}{FP + TP}.
        \end{equation}

    \item \textbf{True Skill Statistic (TSS)}: Balances POD and probability of false detection (POFD):
        \begin{equation}
        TSS = \mathrm{POD} - \frac{FP}{FP + TN}.
        \end{equation}

    \item \textbf{Heidke Skill Score (HSS)}: Measures overall forecast skill relative to random chance using all four entries of the contingency table:
        \begin{equation}
        HSS = \frac{2\,(TP \cdot TN - FP \cdot FN)}{(TP + FN)(FN + TN) + (TP + FP)(FP + TN)}.
        \end{equation}
    
    \item \textbf{Brier Score (BS)}: Mean squared error between predicted probabilities \(f_i\) and observed outcomes \(o_i\), encoded as 1 for flare occurrence and 0 otherwise:
        \begin{equation}
        BS = \frac{1}{N} \sum_{i=1}^{N} (f_i - o_i)^2,
        \end{equation}
        where \(o_i \in \{0,1\}\) is a binary indicator of event occurrence.

    \item \textbf{Brier Skill Score (BSS)}: Measures the improvement of a probabilistic forecast over a reference forecast in terms of Brier Score:
        \begin{equation}
        BSS = 1 - \frac{BS_{\mathrm{model}}}{BS_{\mathrm{ref}}},
        \end{equation}
        where \(BS_{\mathrm{ref}}\) is the Brier Score of a chosen reference forecast. In this study we use the SHARPs-based forecast as the reference when computing the BSS of the RARPs-based model.
\end{itemize}

Most of the presented metrics are domain-specific and capture distinct aspects of forecast quality. Since the models output event probabilities, all contingency-table metrics (POD, FAR, TSS, HSS) require a probability threshold to convert probabilistic forecasts into binary “flare / no flare” decisions. Rather than adopting a single conventional threshold of 0.5, which can be inadequate for rare events, we evaluate these scores over a grid of thresholds from 0.10 to 0.60 in steps of 0.05. This threshold sweep allows us to examine how forecast skill and the balance between hits and false alarms change with the operating point, instead of basing conclusions on one arbitrary cut. ROC-AUC, together with the Brier Score and Brier Skill Score, is computed directly from the continuous predicted probabilities and is used to assess the overall discriminative power and probabilistic calibration of the models, independently of any particular threshold. By probabilistic calibration we mean the agreement between predicted flare probabilities and the observed event frequencies, quantified through the Brier Score and the Brier Skill Score of the RARPs-based model relative to the SHARPs-based reference forecast.

\subsection{Model Training and Evaluation Methodology}

For the forecasting experiments we constructed two synchronized datasets: one based on SHARPs magnetic parameters and one based on RARPs radio embeddings. The SHARPs dataset was built from yearly SHARPs CSV files and the GOES flare event history by merging per-AR, per-time records with flare-activity labels and then removing any rows with missing values in the 18 SHARPs parameters, timing columns, or flare-history fields. The RARPs dataset was derived from the 64-dimensional autoencoder embeddings and synchronized with the SHARPs and flare catalogs via a precomputed matching table linking RARPs filenames to SHARPs records.

We employed a rigorous time-series aware evaluation protocol to ensure realistic assessment of forecasting performance. The dataset was divided into training (2019–2024) and testing (2025) periods, with time-series cross-validation applied to the training set to prevent data leakage. Additionally, we ensured that data from the same active region (AR) never appeared in both training and testing subsets.

To keep the comparison between magnetic and radio predictors transparent and interpretable, this study deliberately employs baseline logistic regression with all hyperparameters left at their default values, except for \texttt{class\_weight}, which was set to \texttt{"balanced"} to address class imbalance via loss weighting rather than resampling or threshold tuning. Two logistic regression models were trained:

\begin{itemize}
    \item \textbf{SHARPs model}: using the standard set of 18 extensive SHARPs magnetic parameters listed in Table~\ref{tab:sharp_params}.
    \item \textbf{RARPs model}: using the 64-dimensional latent embeddings produced by the convolutional autoencoder from the RARPs radio spectra.
\end{itemize}

For the RARPs model, the 64-dimensional latent embedding used as input features is computed from the calibrated Stokes $I$ and $V$ spectra of each active-region patch. During the construction and synchronization of the RARPs products, FITS-header keywords such as \texttt{AR\_NUM}, \texttt{DATE-OBS}, \texttt{TIME-OBS} and \texttt{AZIMUTH} are used solely to associate each radio observation with the corresponding NOAA active region and SHARPs record and to correct for azimuth-dependent gain variations, but they are not included as separate metadata predictors in the forecasting models.

\begin{table}[h!]
\centering
\caption{SHARPs magnetic parameters used as predictors in the logistic regression model.}
\label{tab:sharp_params}
\scriptsize
\begin{tabular}{p{0.3\columnwidth} >{\centering\arraybackslash}p{0.6\columnwidth}}
\toprule
\textbf{Name} & \textbf{Description} \\
\midrule
USFLUX   & Total unsigned magnetic flux \\
MEANGAM  & Mean inclination angle of the field vector \\
MEANGBT  & Mean total field gradient \\
MEANGBZ  & Mean vertical field gradient \\
MEANGBH  & Mean horizontal field gradient \\
MEANJZD  & Mean vertical current density \\
TOTUSJZ  & Total unsigned vertical current \\
MEANALP  & Mean force–free twist parameter \\
MEANJZH  & Mean current helicity \\
TOTUSJH  & Total unsigned current helicity \\
ABSNJZH  & Absolute value of net current helicity \\
SAVNCPP  & Sum of absolute net currents per polarity \\
MEANPOT  & Mean photospheric free magnetic energy density \\
TOTPOT   & Total photospheric free magnetic energy density \\
MEANSHR  & Mean shear angle of the field \\
SHRGT45  & Fraction of pixels with shear angle $>45^\circ$ \\
AREA\_ACR & Deprojected AR area \\
R\_VALUE & Schrijver $R$ parameter (flux near strong PILs) \\
\bottomrule
\end{tabular}
\end{table}

All preprocessing, autoencoder training, and logistic regression experiments, including the exact hyperparameters (learning rate, optimizer, batch size, loss function used prior to the Brier score analysis, and all convolutional kernel sizes), are implemented in an open-source repository\footnote{\protect\url{https://github.com/SpbfSAO/RARPs_CaseStudy}}. For convenience, the main hyperparameters of the autoencoder and logistic regression models are also summarized in Appendix~B (Tables~\ref{tab:all-hparams}).

The flare‑prediction experiment presented here should be viewed as a baseline demonstration of the information content in the RARPs and SHARPs predictors, rather than as a fully optimized operational forecasting system. The reported skill scores are conservative lower bounds that more advanced architectures are likely to surpass.

\subsection{Performance Summary: SHARPs vs. RARPs}

Across all experiments, the SHARPs model demonstrated higher discriminative performance than the RARPs model. For C-class flare forecasts, the SHARPs models achieved ROC-AUC values of 0.711 at 24h and 0.682 at 48h, compared with 0.630 and 0.602 for RARPs. For M+ flare forecasts, the difference was even larger: SHARPs reached 0.805 (24h) and 0.790 (48h), while RARPs produced values of 0.637 and 0.648. These results establish a consistent advantage of SHARPs over RARPs in terms of separation between flare and non-flare cases.

\renewcommand{\arraystretch}{1.3}
\begin{table*}[ht]
\centering
\caption{Summary of metrics for the SHARPs and RARPs logistic regression models, evaluated for C–class and $\ge$M–class flare targets over 0–24~h and 24–48~h prediction windows at a probability threshold $p_{\mathrm{thr}} = 0.50$.}
\label{tab:summary_all}
\scriptsize
\begin{tabularx}{\textwidth}{@{}l c c 
                                >{\centering\arraybackslash}X
                                >{\centering\arraybackslash}X
                                >{\centering\arraybackslash}X
                                >{\centering\arraybackslash}X
                                >{\centering\arraybackslash}X
                                >{\centering\arraybackslash}X
                                >{\centering\arraybackslash}X
                                @{}}
\toprule
\textbf{Target} & \textbf{Window} & \textbf{Model} &
AUC & Brier & BSS &
TSS & HSS & FAR & POD \\
\midrule
\multirow{4}{*}{C-class} 
  & \multirow{2}{*}{0--24 h} 
    & SHARPs & 0.711 & 0.231 & \phantom{$-$}0.000 & 0.336 & 0.278 & 0.563 & 0.758 \\
  &                                   & RARPs & 0.630 & 0.230 & \phantom{$-$}0.004 & 0.193 & 0.186 & 0.581 & 0.481 \\
  & \multirow{2}{*}{24--48 h} 
    & SHARPs & 0.682 & 0.240 & \phantom{$-$}0.000 & 0.295 & 0.232 & 0.614 & 0.722 \\
  &                                   & RARPs & 0.602 & 0.233 & \phantom{$-$}0.029& 0.166 & 0.151 & 0.637 & 0.477 \\
\midrule
\multirow{4}{*}{$\ge$M-class} 
  & \multirow{2}{*}{0--24 h} 
    & SHARPs & 0.805 & 0.234 & \phantom{$-$}0.000 & 0.528 & 0.168 & 0.843 & 0.883 \\
  &                                   & RARPs & 0.637 & 0.187 & \phantom{$-$}0.200 & 0.250 & 0.117 & 0.863 & 0.474 \\
  & \multirow{2}{*}{24--48 h} 
    & SHARPs & 0.790 & 0.235 & \phantom{$-$}0.000 & 0.464 & 0.134 & 0.869 & 0.816 \\
  &                                   & RARPs & 0.648 & 0.193 & \phantom{$-$}0.179 & 0.250 & 0.099 & 0.884 & 0.493 \\
\bottomrule
\end{tabularx}
\end{table*}
The SHARPs model also demonstrated superior forecasting performance, with its highest TSS and HSS values achieved at probability thresholds between 0.5 and 0.6. This reflects a favourable balance between the POD and POFD entering the TSS definition, while the Heidke Skill Score shows consistent behaviour and confirms the same optimal threshold range. A notable trend is that as the threshold decreases, the POD sharply increases, but this comes at the cost of a higher false alarm ratio (FAR) and a lower overall skill score, highlighting the inherent trade-off between sensitivity and reliability.

For C-class flares, both TSS and HSS peak for the 24-hour and 48-hour predictions at a threshold of 0.5. For the more energetic M+ flares, the TSS (True Skill Statistic) is significantly higher, peaking at 0.5 for the 24-hour forecast and at 0.6 for the 48-hour forecast. However, the HSS (Heidke Skill Score) is lower for M+ flares, a result of the high False Alarm Ratio (FAR) observed, which reflects the model's strong focus on maximizing detection (high POD) at the expense of reliability. This contrast highlights the better discrimination capability (TSS) but lower overall reliability (HSS) for the rarer, more energetic events. Considering these skill scores as the primary metrics, the best SHARPs model results are:

\begin{itemize}
    \item \textit{C-class flares}:
    \begin{itemize}
        \item \textit{24-hour prediction}: TSS = 0.336, HSS = 0.278 (Threshold 0.5, POD = 0.76, FAR = 0.56)
        \item \textit{48-hour prediction}: TSS = 0.295, HSS = 0.232 (Threshold 0.5, POD = 0.72, FAR = 0.61)
    \end{itemize}
    \item \textit{M+ flares}:
    \begin{itemize}
        \item \textit{24-hour prediction}: TSS = 0.528, HSS = 0.168 (Threshold 0.5, POD = 0.88, FAR = 0.84)
        \item \textit{48-hour prediction}: TSS = 0.520, HSS = 0.201 (Threshold 0.6, POD = 0.76, FAR = 0.83)
    \end{itemize}
\end{itemize}

The RARPs model, while showing a lower overall skill compared to SHARPs, still demonstrated some capability in flare discrimination. Its features allow for a degree of forecasting skill, as evidenced by its positive TSS values. However, its performance is limited by a lower POD. The trend is clear: even as the POD improves with lower thresholds, the alarm rates become so high that the model's utility is compromised. It is important to note that the high false alarm ratio is a shared challenge with the SHARPs model, particularly for M+ flares. The overall low TSS values for the RARPs model indicate that their features are not yet sufficient for a reliable stand-alone forecasting system, but they show clear potential. Prioritizing TSS and HSS as key metrics, the best results for the RARPs model were:

\begin{itemize}
    \item \textit{C-class flares}:
    \begin{itemize}
        \item \textit{24-hour prediction}: TSS = 0.193, HSS = 0.186 (Threshold 0.5, POD = 0.48, FAR = 0.58)
        \item \textit{48-hour prediction}: TSS = 0.166, HSS = 0.151 (Threshold 0.5, POD = 0.48, FAR = 0.64)
    \end{itemize}
    \item \textit{M+ flares}:
    \begin{itemize}
        \item \textit{24-hour prediction}: TSS = 0.261, HSS = 0.117 (Threshold 0.55, POD = 0.43, FAR = 0.84)
        \item \textit{48-hour prediction}: TSS = 0.261, HSS = 0.083 (Threshold 0.45, POD = 0.59, FAR = 0.90)
    \end{itemize}
\end{itemize}

From a probabilistic perspective, Brier Scores and Brier Skill Scores (BSS) reveal a mixed picture. For C-class flares, the RARPs model are slightly better calibrated than SHARPs, with Brier Scores of 0.230 (24 h) and 0.233 (48 h) compared with 0.232 and 0.240 for SHARPs, corresponding to small positive skill scores \(BSS_{\mathrm{RARPs|SHARPs}} \approx -0.13\) and \(-0.05\). For M+ flares the RARPs-based forecasts are better calibrated: RATAN achieves Brier Scores of 0.187 (24 h) and 0.193 (48 h) versus 0.234 and 0.235 for SHARPs, yielding positive \(BSS_{\mathrm{RARPs|SHARPs}} \approx 0.20\) and \(0.18\). This indicates that, although the radio-based model has weaker discriminative skill than SHARPs, its predicted probabilities for major (M+) flares are on average closer to the observed event frequencies.

\subsection{Evolutionary curve}

The temporal coverage and standardized format of the RARPs database enables systematic studies of AR evolution across their lifetimes. An example of such evolutionary curves for NOAA AR~12371, showing the time evolution of peak amplitude and integrated area in Stokes $I$ and $V$, is presented in Fig.~\ref{fig:evolution_curves_I} and Fig.~\ref{fig:evolution_curves_V} in Appendix~\ref{app:evolution}. This capability addresses fundamental questions about AR development, magnetic field evolution, and the buildup to eruptive events \cite{knyazeva2010topology, makarenko2012magnetic, knyazeva2015comparison}. RARPs enable researchers to track how radio emission characteristics change as ARs emerge, develop, and decay. The database structure supports both individual case studies of well-observed regions and statistical analyses of AR populations.

Time series should be interpreted with caution. Given the scale of RATAN-600, deriving absolute values for AR characteristics and embedded microwave sources is challenging due to variations in receiver performance and antenna beam pattern. Consequently, AR time series are best regarded as relative measures, with comparisons most reliable when restricted to the same observation azimuths. This reduces the available sample but minimizes beam-related uncertainties. With care, multi-azimuth data can also be used to probe the hourly evolution of ARs and their sources. Several studies have reported parameter changes both before and after flares. Although the likelihood of capturing a flare itself is low, given their rarity, it remains possible to trace AR and source evolution at least at the level of overall trends, if not through absolute measurements.

ARs are of particular interest in both pre-flare and post-flare stages. Pre-flare studies of AR fluxes and microwave brightness provide insight into the mechanisms that drive flares. Two key aspects can be distinguished: the amount of stored energy sufficient to power a flare, and the trigger that releases it. Numerous studies have addressed such mechanisms. As a proxy for stored energy, one may use the microwave flux or AR brightness, since large ARs are typically flare-productive. Potential triggers include the emergence of new magnetic flux or changes in emission configuration, manifested as variations in microwave flux at one or several frequencies.

The post-flare stage is particularly valuable for probing structural changes in ARs. Multi-frequency time series of microwave flux can indicate not only which source within the AR was involved, but also reveal modifications in its vertical structure.

Of special interest are ARs that persist not only for several days during their passage across the visible solar disk (approx. 12-14 days, from which 8–9 days within the well-observed  zone of the disk \(\pm 60^\circ\) from  central meridian distance), but also for one or even multiple solar rotations. Such long-lived ARs provide a unique opportunity to study their extended temporal evolution.

\section{Discussion}

The automated extraction and standardization pipeline transforms decades of RATAN-600 observations into analysis-ready data products. Robust outlier handling, coordinate alignment, flexible windowing, and metadata integration ensure high-quality outputs compatible with existing astronomical tools. The library simplifies both individual research projects and large-scale statistical studies.

We have also constructed the first comprehensive database of standardized solar AR radio signatures (RARPs), addressing a critical gap in multi-wavelength solar observations. Combined with enhanced \texttt{RATANSunPy} capabilities, RARPs establishes a new standard for radio data accessibility, enabling reproducible analyses and facilitating cross-disciplinary studies.

Key strengths of this work include the first standardized radio active-region database RARPs as a direct counterpart to SHARPs, a fully reproducible RATANSunPy v2 processing pipeline, multi-frequency 3–18 GHz coverage not available in SHARPs, a long 2009–2025 temporal baseline, and radio features that complement photospheric magnetic parameters for space-weather–oriented flare forecasting. At the same time, the flare forecasting experiments in this study are restricted to the 2019–2025 interval, corresponding primarily to the ascending phase of Solar Cycle 25, so the reported predictive performance should be interpreted as specific to this phase.

Our predictive analysis demonstrates that radio-based models provide information complementary to photospheric magnetic field parameters. While SHARPs-based forecasts achieve superior discrimination of C-class and M+ flare events, RARPs-based models provide comparatively better probabilistic calibration for M+ flares, highlighting the value of multimodal approaches that integrate both data types. However, the robustness of these performance estimates is further constrained by the limited number of flaring regions in some prediction windows and by high false-alarm rates in both SHARPs- and RARPs-based models. 

A fundamental distinction between the two modalities lies in feature extraction. In the photospheric domain, the use of SHARP parameters is the well-established practice; these engineered features are designed to capture specific physical properties of the magnetic field (e.g., non-potentiality), providing a physics-based interpretability to the forecasts.
Conversely, for solar radio emission, a widely accepted set of standardized physical parameters comparable to SHARPs does not yet exist. 

While historical heuristics, such as the Tanaka-Enomé criterion \cite{tanaka1975microwave}, link spectral indices to flaring probability, they are often difficult to generalize automatically across large, diverse datasets. To address this, we opted for a data-driven representation learning approach using a convolutional autoencoder. Unlike manual feature engineering, this method allows us to capture the latent information content of the multi-frequency spectra—including complex spectral-spatial dependencies—without being constrained by the limitations of specific, pre-defined heuristic models. 

During analysis, we observed that the azimuth of RATAN-600 scans introduces systematic changes in signal amplitude, evident in the AR evolution curves. These azimuth-dependent variations were intentionally not corrected in the logistic regression pipeline to maintain a simple baseline analysis, meaning that the present results likely underestimate the achievable predictive skill of RARPs-based models, and roughly evaluate the predictive potential of RARPs-based models, so the resulting time series should be treated primarily as relative measures, with the most reliable comparisons obtained for observations taken at the same or very similar azimuths. In particular, azimuth-dependent amplitude changes limit the direct comparability of RATAN active-region patches across different days and scan geometries, even when the underlying solar region is the same. In future work, these effects could be mitigated by applying more detailed azimuth- and frequency-dependent gain corrections, incorporating azimuth as an explicit covariate, or employing domain-adaptation schemes.

The predictive capability of the radio-based model can be plausibly interpreted through the detection of pre-flare spectral features historically described as ‘Peculiar Sources’ (PS) or Neutral Line Sources (NLS) in RATAN-600 observations \cite{abramov2015dynamics, peterova2021increased}. In many documented cases, these compact sources appear to emerge approximately 1–3 days before major eruptive events, serving as potential early-stage precursors rather than typical active region components such as sunspot-associated or plage-related sources. Observationally, they tend to exhibit a noticeable increase in flux density at shorter wavelengths (2–4 cm) and elevated brightness temperatures that can reach up to ~10 MK, according to published estimates. Their physical origin is commonly attributed to gyrosynchrotron emission from mildly relativistic electrons, likely accelerated in current sheets above magnetic neutral lines or regions of emerging flux.

While SHARPs parameters quantify photospheric boundary conditions (for example, magnetic gradients), the radio signatures arguably probe coronal volumes where magnetic free energy is accumulated and later released. In this context, the autoencoder’s ability to extract latent features from spectral morphology suggests that it may be identifying manifestations of spectral hardening and polarization asymmetries which have been associated in previous studies with the gradual build-up of electric currents in the corona.

The observed advantage of RARPs-based models in Brier Score performance for M-class and above flares therefore implies, although not conclusively, that radio diagnostics may help calibrate probabilistic forecasts by encoding information related to cumulative energy buildup, whereas magnetic parameters appear to characterize conditions closer to the instability onset. This distinction points toward a complementary — rather than redundant — relationship between the two modalities for multimodal flare prediction.

Analysis of the Stokes $V$ channel indicates higher sensitivity to noise and small-scale fluctuations, resulting in noticeably lower reconstruction fidelity than for Stokes $I$, and therefore greater uncertainty in interpreting circular polarization features. Future improvements may include polarization-specific preprocessing (denoising, adaptive normalization, or outlier mitigation) and alternative autoencoder architectures with deeper layers, residual connections, or attention mechanisms. Joint training on both $I$ and $V$ channels could leverage correlations to improve reconstruction accuracy. The MAD-based normalization approach and time-series aware evaluation protocol provide templates for other 2D spectral datasets and space weather prediction frameworks. It should also be noted that RARPs, as used here, are not explicitly corrected for long-term instrument evolution.

Planned enhancements for RARPs and the library include integration of additional frequency ranges (notably the 1–3~GHz RATAN-600 spectrometer), improved polarization processing, and development of specialized deep learning architectures for spectral feature extraction. Expanding AR analysis to track temporal evolution, magnetic field dynamics, and multi-wavelength spectral properties will further improve flare forecasting and the understanding of underlying physical mechanisms.

\section{Conclusions}

This work presents \texttt{RATANSunPy} v2 and the RARPs database, establishing a comprehensive framework for solar radio astronomy research. The automated processing pipeline transforms decades of RATAN-600 observations into standardized, analysis-ready data products, while RARPs provides the first comprehensive catalog of solar AR radio signatures spanning from 2009 to 2025.

Our machine learning analysis demonstrates that radio-based models complement traditional magnetic field approaches for solar flare forecasting: SHARPs parameters achieve consistently higher discriminative performance, whereas for M-class and above flares the RARPs-derived features attain smaller Brier Scores and positive Brier Skill Scores relative to the SHARPs-based reference, pointing to more accurate probability calibration in that specific regime, highlighting the potential value of multimodal prediction systems. The ground-based nature of RATAN-600 provides crucial operational redundancy to space-based instruments for space weather monitoring.

Key challenges remain intensity and circular polarization processing, guiding future development priorities. Planned enhancements include integration of 1-3 GHz data, improved polarization analysis, and specialized deep learning architectures for spectral feature extraction.

By delivering standardized radio data products and flexible analysis tools, this project fundamentally improves accessibility to radio solar observations. RATANSunPy and RARPs provide essential infrastructure for advancing solar activity research and space weather prediction capabilities, serving as a template for modernizing other long-term observational datasets through contemporary data science approaches.

\noindent\section*{Declaration of competing interest}
The authors declare that they have no known competing financial interests or personal relationships that could have appeared
to influence the work reported in this paper.
\section*{Data availability}
The data for which this project is designed to process is publicly available. Raw data from RATAN-600 available from the site of St. Petersburg branch of the SAO  \footnote{\protect\url{http://spbf.sao.ru/data/ratan/}}. Solar Region Summary available from Space Weather Prediction Center Of National Oceanic And Atmospheric Administration via http or ftp \footnote{\protect\url{https://www.swpc.noaa.gov/content/data-access}}.

\section*{Acknowledgements}
Observations with the SAO RAS telescopes are supported by the Ministry of Science and Higher Education of the Russian Federation. The renovation of telescope equipment is currently provided within the national project "Science and Universities".

The authors would like to thank the colleagues who participated in the development and testing of the information system \href{http://www.spbf.sao.ru/prognoz/}{Prognoz}, on the basis of which this package was implemented, as well as the senior researcher of the St. Petersburg branch of the SAO RAS Bogod V.M. for the support of this work. 

We acknowledge the essential contributions of large Python packages, such as \texttt{NumPy} \footnote{\url{https://numpy.org/}}, \texttt{SciPy} \footnote{\url{https://scipy.org/}}, \texttt{AstroPy} \footnote{\url{https://www.astropy.org/}} and \texttt{Matplotlib} \footnote{\url{https://matplotlib.org/}}, which are critical for the functionality of the RATAN-600 data processing pipeline, providing key tools for numerical analysis and visualization.

\section*{Funding}
The research was carried out under the Russian Science Foundation grant No.24-21-00476, \href{https://rscf.ru/project/24-21-00476/}{Link to the project card.}

\clearpage
\onecolumn
\appendix

\section*{Appendix A: Example of AR evolutionary curves}
\label{app:evolution}

\renewcommand{\thefigure}{A.\arabic{figure}}
\setcounter{figure}{0}

\renewcommand{\thetable}{A.\arabic{table}}
\setcounter{table}{0}

\begin{figure*}[ht]
    \centering

    \begin{subfigure}[b]{\textwidth}
        \centering
        \includegraphics[width=\textwidth]{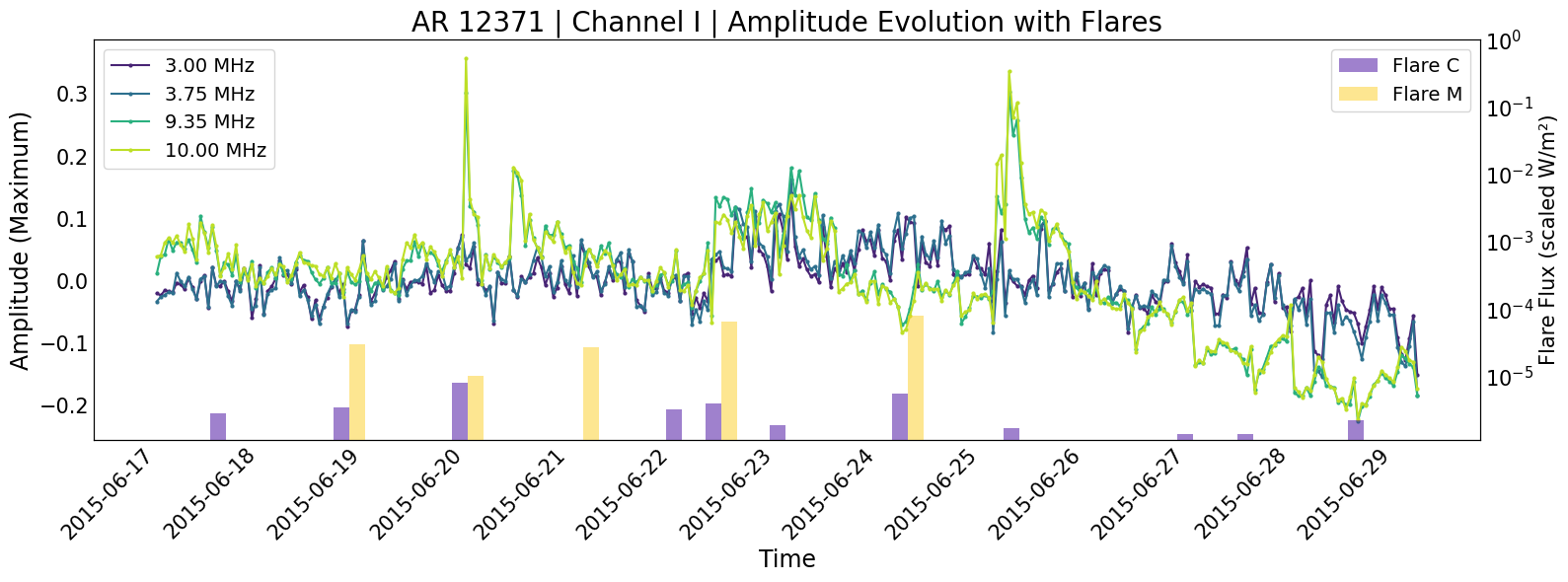}
        \caption{I amplitude evolution}
        \label{fig:evol_i_amp}
    \end{subfigure}

    \vspace{0.8em}

    \begin{subfigure}[b]{\textwidth}
        \centering
        \includegraphics[width=\textwidth]{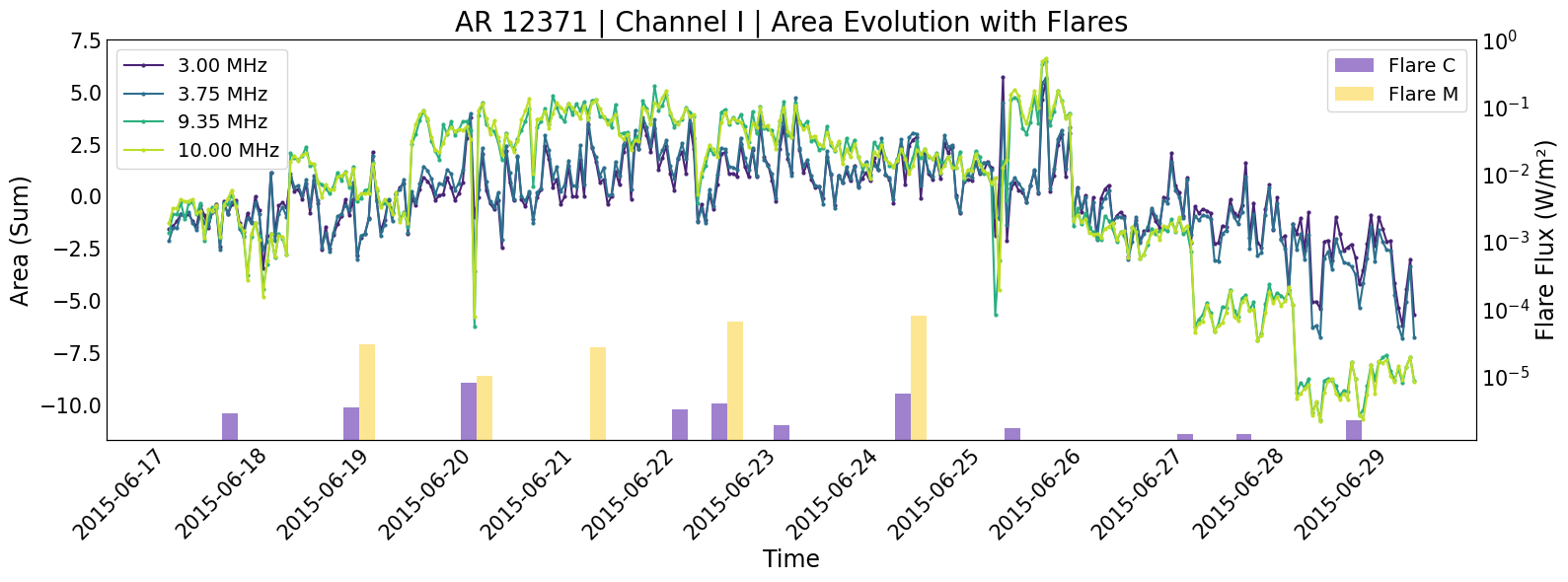}
        \caption{I area evolution}
        \label{fig:evol_i_area}
    \end{subfigure}

    \caption{Evolutionary curves for NOAA AR~12371 in intensity (Stokes $I$). Panels (a) and (b) show the time evolution of peak amplitude and integrated area at selected RATAN-600 frequencies.}
    \label{fig:evolution_curves_I}
\end{figure*}

\clearpage

\begin{figure*}[ht]
    \centering

    \begin{subfigure}[b]{\textwidth}
        \centering
        \includegraphics[width=\textwidth]{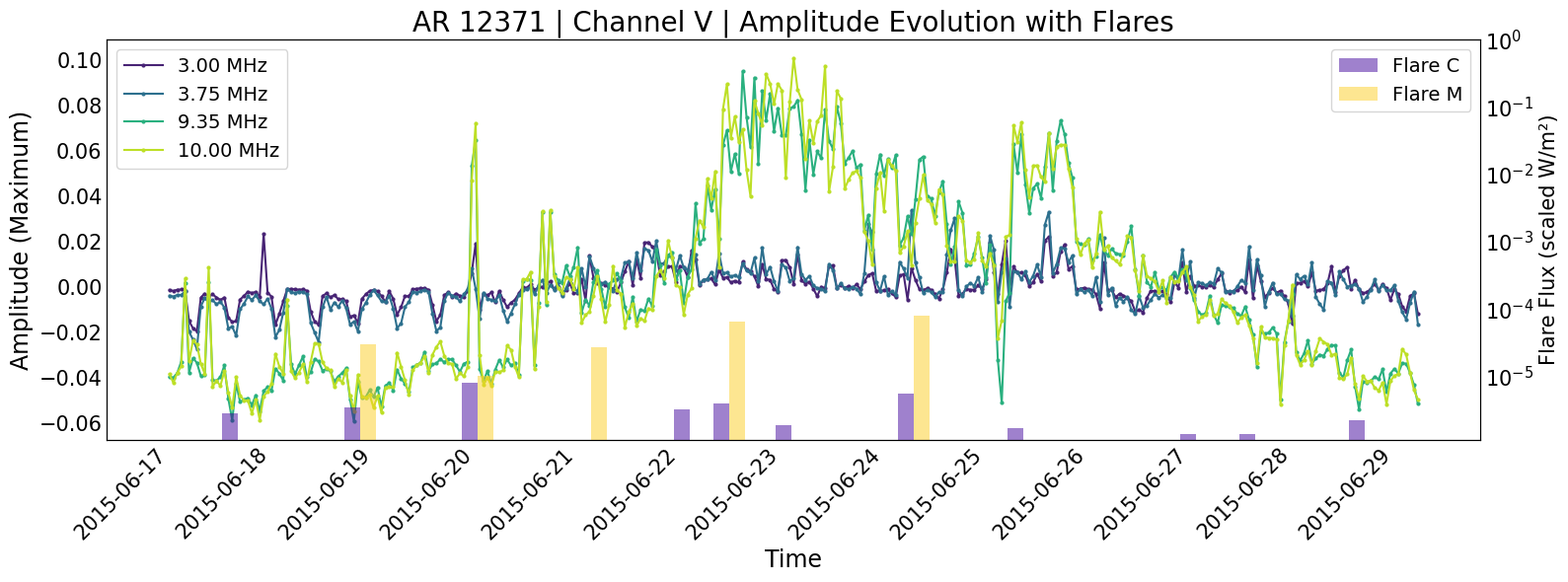}
        \caption{V amplitude evolution}
        \label{fig:evol_v_amp}
    \end{subfigure}

    \vspace{0.8em}

    \begin{subfigure}[b]{\textwidth}
        \centering
        \includegraphics[width=\textwidth]{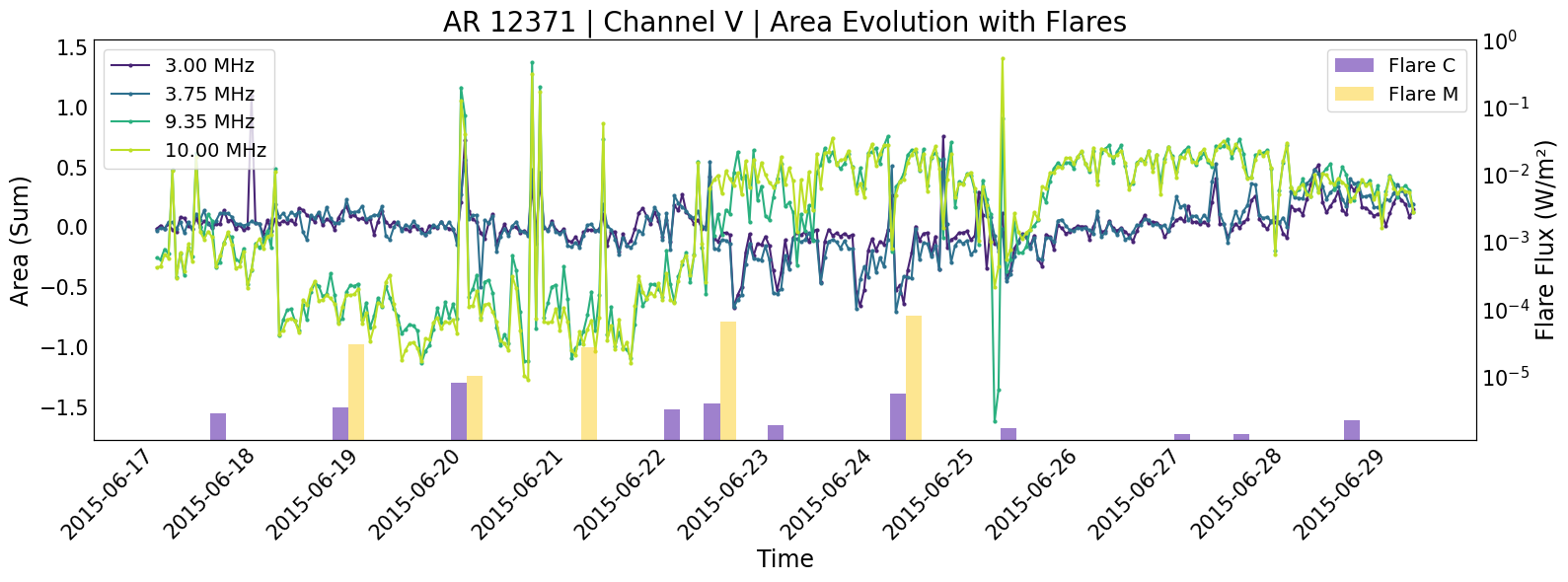}
        \caption{V area evolution}
        \label{fig:evol_v_area}
    \end{subfigure}

    \caption{Evolutionary curves for NOAA AR~12371 in circular polarization (Stokes $V$). Panels (c) and (d) show the time evolution of peak amplitude and integrated area at selected RATAN-600 frequencies.}
    \label{fig:evolution_curves_V}
\end{figure*}

\clearpage

\section*{Appendix B: Forecasting Pipeline summary and Hyperparameters}
\label{app:pipeline_hyperparams}

\renewcommand{\thefigure}{B.\arabic{figure}}
\setcounter{figure}{0}

\renewcommand{\thetable}{B.\arabic{table}}
\setcounter{table}{0}

\begin{table*}[h!]
\centering
\caption{Forecasting pipeline: data sources, alignment, preprocessing, feature extraction, and evaluation.}
\label{tab:forecast_pipeline_summary}
\scriptsize 
\begin{tabularx}{\textwidth}{@{}l X X@{}}
\toprule
\textbf{Stage} & \textbf{Inputs} & \textbf{Description} \\
\midrule

Data sources &
RARPs AR spectra (RATAN-600); SHARPs magnetic parameters; GOES flare catalogs &
RARPs provide 2D Stokes~I/V patches and metadata (NOAA AR, time, azimuth). SHARPs supply 18 standard magnetic indices. GOES lists flares with start times, classes, and NOAA AR numbers. \\
\midrule

Temporal and AR alignment &
RARPs filenames; NOAA$\rightarrow$HARP mapping; SHARPs time series; GOES events &
Each RARPs spectrum is treated as a separate sample. NOAA ARs are mapped to HARP numbers, and SHARPs patches are queried at the exact RATAN time. GOES flares are matched by NOAA AR and spectrum time using 0--24\,h and 24--48\,h prediction windows. \\
\midrule

Radio preprocessing and normalization &
Calibrated RARPs Stokes~I/V maps &
NaN values are mapped to zero. Mean Stokes~I/V per patch is modeled as a function of azimuth and subtracted. Heavy-tailed artefacts are mitigated via MAD normalization and clipping, followed by global scaling. \\
\midrule

Feature extraction &
Preprocessed RARPs patches; SHARPs headers &
A convolutional autoencoder compresses each 2D RARPs patch into a 64-dimensional latent embedding. SHARPs features are the 18 extensive magnetic parameters. \\
\midrule

Label construction &
Aligned RARPs--SHARPs samples; GOES flare list &
Binary targets for C (C1.0--C9.9) and M+ (M1.0+, incl. X) events are defined in 0--24\,h and 24--48\,h windows after each spectrum time. \\
\midrule

Classifier training and evaluation &
RARPs embeddings; SHARPs features; train/test split &
Two logistic regression baselines are trained (SHARPs vs. RARPs). Temporal split: 2019--2024 training; 2025 testing; no AR overlap between splits. Metrics: ROC--AUC, TSS, HSS, Brier Score, Brier Skill Score. \\
\bottomrule
\end{tabularx}

\end{table*}

\begin{table*}[ht]
\centering
\caption{Hyperparameters of the autoencoder and logistic regression models.}
\label{tab:all-hparams}
\scriptsize

\begin{minipage}[t]{0.48\textwidth}
\centering
\textbf{Autoencoder}

\vspace{2mm}
\begin{tabular}{@{}l l@{}}
\toprule
Parameter & Value \\
\midrule
Input image height & 75 \\
Input image width & 101 \\
Latent dimension & 64 \\
Epochs & 20 \\
Batch size & 32 \\
Learning rate & 0.001 \\
Optimizer & Adam \\
Loss function & Mean squared error \\
Convolution kernel size & $3 \times 3$ \\
Pooling kernel/stride & $2 \times 2$ \\
Activation function & ReLU \\
\end{tabular}
\end{minipage}
\hfill
\begin{minipage}[t]{0.48\textwidth}
\centering
\textbf{Logistic regression}

\vspace{2mm}
\begin{tabular}{@{}l l@{}}
\toprule
Parameter & Value \\
\midrule
Class weight & balanced \\
Penalty & $\ell_2$ \\
Inverse of regularization strength $C$ & 1.0 \\
Solver & lbfgs \\
Maximum iterations & 100 \\
Random state & 17 \\
Input scaling & StandardScaler \\
CV scheme & TimeSeriesSplit (2 folds) \\
Output type & Class probabilities (predict\_proba) \\
\end{tabular}
\end{minipage}

\end{table*}

\twocolumn

\bibliographystyle{elsarticle-harv} 
\bibliography{ratan}

\end{document}